\PassOptionsToPackage{bookmarks=false}{hyperref}
\documentclass[]{pasj02} 
\usepackage[switch,mathlines]{lineno} 

\jyear{2026}
\Received{}
\Accepted{}

\usepackage{tikz}
\usetikzlibrary{arrows.meta,calc,shapes.geometric,decorations.pathmorphing,decorations.pathreplacing}
\usepackage{graphicx}
\usepackage[scaled]{helvet} 
\usepackage{setspace}

\usepackage{url,lscape}

\definecolor{xlinkcolor}{rgb}{0,0,1}
\definecolor{xurlcolor}{cmyk}{0.78,0.17,0.09,0} 
\makeatletter
\let\old@ssect\@ssect 
\makeatother

\usepackage{natbib,amsmath}
\usepackage[pdfpagelabels=false]{hyperref}
\hypersetup{colorlinks=true,linkcolor=red,citecolor=xlinkcolor,urlcolor=xurlcolor}

\makeatletter
\def\@ssect#1#2#3#4#5#6{%
  \NR@gettitle{#6}
  \old@ssect{#1}{#2}{#3}{#4}{#5}{#6}
}
\makeatother

\newcommand{\rqq}{\textquotedblright}
\newcommand{\lqq}{\textquotedblleft}

\begin{document} 

\title{Assessing the Impact of Source Confusion for GREX-PLUS based on Deep JWST NIRCam Imaging}

\author{
Yoshiaki \textsc{Ono},\altaffilmark{1}\altemailmark\orcid{0000-0001-9011-7605} \email{ono@icrr.u-tokyo.ac.jp} 
Akio K. \textsc{Inoue},\altaffilmark{2,3}\orcid{0000-0002-7779-8677} 
Yuma \textsc{Sugahara},\altaffilmark{2,3}\orcid{0000-0001-6958-7856}
Takeshi \textsc{Hashigaya},\altaffilmark{4}
Fumihide \textsc{Iwamuro},\altaffilmark{4}\orcid{0000-0002-7419-2629}
Taiki \textsc{Bessho},\altaffilmark{5}\orcid{0000-0002-2179-7560}
Yuji \textsc{Ikeda},\altaffilmark{5,6}
Matthew L. N. \textsc{Ashby},\altaffilmark{7}\orcid{0000-0002-3993-0745}
Yuichi \textsc{Harikane},\altaffilmark{1}\orcid{0000-0002-6047-430X}
Jarron \textsc{Leisenring},\altaffilmark{8}\orcid{0000-0002-0834-6140}
Takao \textsc{Nakagawa},\altaffilmark{9,10}\orcid{0000-0002-6660-9375}
and 
Howard A. \textsc{Smith}\altaffilmark{7}
}
\altaffiltext{1}{Institute for Cosmic Ray Research, The University of Tokyo, 5-1-5 Kashiwanoha, Kashiwa, Chiba 277-8582, Japan}
\altaffiltext{2}{Waseda Research Institute for Science and Engineering, Faculty of Science and Engineering, Waseda University, 3-4-1 Okubo, Shinjuku, Tokyo 169-8555, Japan}
\altaffiltext{3}{Department of Pure and Applied Physics, School of Advanced Science and Engineering, Faculty of Science and Engineering, Waseda University, 3-4-1 Okubo, Shinjuku, Tokyo 169-8555, Japan}
\altaffiltext{4}{Department of Astronomy, Kyoto University, Kitashirakawa, Sakyo-ku, Kyoto 606-8502, Japan}
\altaffiltext{5}{Photocross., Co. Ltd., 819-1 BLDG D, Iwakura-chuzaichicho, Sakyo-ku, Kyoto, 606-0021, Japan} 
\altaffiltext{6}{Laboratory of Infrared High-resolving Spectroscopy, Koyama Astronomical Observatory, Kyoto-Sangyo University, Motoyama, Kamigamo, Kita-ku, Kyoto 603-8555, Japan}
\altaffiltext{7}{Center for Astrophysics | Harvard {\&} Smithsonian, 60 Garden St., Cambridge, MA 02138-1516, USA}
\altaffiltext{8}{Steward Observatory, University of Arizona, Tucson, AZ 85721, USA}
\altaffiltext{9}{Institute of Space and Astronautical Science, Japan Aerospace Exploration Agency, 3-1-1 Yoshinodai, Chuo-ku, Sagamihara, Kanagawa 252-5210, Japan}
\altaffiltext{10}{Advanced Research Laboratories, Tokyo City University, 1-28-1 Tamazutsumi, Setagaya-ku, Tokyo 158-8557, Japan}




\maketitle

\begin{abstract}
\begin{spacing}{1.03}
\normalfont\rmfamily
We investigate the effects of source confusion 
expected in observations with Galaxy Reionization EXplorer and PLanetary Universe Spectrometer (GREX-PLUS),
a JAXA L-class mission candidate for a space infrared telescope 
equipped with a wide-field infrared camera covering $2$--$8\,\mu$m with a field of view of $0.50$~deg$^2$.
Focusing on the deep imaging band near $4\,\mu$m,
we calculate the GREX-PLUS point spread function (PSF) and ghost based on the latest optical design, 
and consider two representative imaging performance cases with PSF FWHM values of $0.9$ and $1.2$~arcsec. 
We then construct simulated GREX-PLUS images at different depths 
by convolving JWST NIRCam imaging data
from JADES, GLASS, CEERS, and COSMOS-Web with the PSF$+$ghost kernel.
Comparing the limiting magnitudes estimated from random aperture photometry 
using the same aperture sizes,
we find that the simulated GREX-PLUS images are shallower than the original JWST images, 
with the deviation becoming more pronounced 
as the original JWST images become deeper.
This likely reflects unresolved faint sources and extended PSF$+$ghost wings from bright sources, 
which elevate background fluctuations in blank regions. 
Nevertheless, the limiting magnitudes derived from blank sky aperture fluctuations 
continue to improve with increasing integration time down to $\simeq27$~mag, 
without showing a clear plateau at depths comparable to the planned GREX-PLUS deep survey,
although the improvement becomes progressively less efficient toward longer integrations.
Based on Monte Carlo simulations, 
we estimate detection completeness and correct the number counts
for magnitude bias and incompleteness,  
finding that confusion-induced blending can reduce the completeness 
even at magnitudes well above the nominal $5\sigma$ depth.
The completeness-corrected number counts agree well with 
the JWST-based number counts down to around the detection limit. 
Overall, our results suggest that statistical studies of faint galaxies remain feasible for GREX-PLUS;
however, survey planning should account for less efficient depth improvement toward longer integrations 
as source confusion effects become more important.
\end{spacing}
\end{abstract}


\section{Introduction}
\label{sec:Introduction}

\hspace{0.8em}
Galaxy Reionization EXplorer and PLanetary Universe Spectrometer
(GREX-PLUS; \citealt{2022SPIE12180E..1II}; \citealt{2023arXiv230408104G}; \citealt{2024SPIE13092E..0YI})
is a mission candidate for a space infrared telescope with a $1.0$~m primary mirror 
and a focal ratio of $f/7.38$,
proposed as a JAXA strategic L-class mission to be launched in the 2030s.
One of its primary science objectives is to conduct a wide-area deep near-infrared imaging survey
using the wide-field camera (WFC), which provides a large field of view of $0.50$~deg$^2$ 
with wavelength coverage from $2\,\mu$m to $8\,\mu$m,
to systematically identify rare bright high-redshift sources, 
providing a foundation for studies of galaxy formation and evolution.

When planning deep imaging surveys at long wavelengths, 
it is crucial to quantify the impact of source confusion  
(\citealt{1974ApJ...188..279C};
\citealt{1989ApJ...344...35F};
\citealt{1990LIACo..29..117H}; 
\citealt{1991A&AS...89..285F};
\citealt{1995SSRv...74...17R};
\citealt{2001AJ....121.1207H}; 
\citealt{2001MNRAS.325.1241V}; 
\citealt{2003ApJ...585..617D}; 
\citealt{2003PASJ...55..717J}; 
\citealt{2004ApJ...604...40T}; 
\citealt{2004MNRAS.352..493N}; 
\citealt{2004ApJS..154...93D}; 
\citealt{2005A&A...430..343K};
\citealt{2006MNRAS.369..281J};
\citealt{2010A&A...518L...5N};
\citealt{2013A&A...553A.132M}; 
\citealt{2024A&A...692A..52B}).
As illustrated in Figure~1 of \cite{2010A&A...518L...5N},
source confusion becomes increasingly important at long wavelengths,
because the diffraction-limited angular resolution degrades in proportion to wavelength
and the surface density of faint distant sources on the sky is high.
As a consequence, an increasing fraction of sources becomes unresolved, 
and fluctuations in their combined emission contribute to the background.
In this paper, we refer to this background fluctuation component, 
arising from the superposition of unresolved sources, as confusion noise.
Because distant galaxies are distributed approximately isotropically on large scales,
there is no straightforward way to subtract the confusion noise contribution from images.
Moreover, unlike detector noise and photon noise, 
confusion noise cannot be reduced simply by increasing the integration time.
Separately, finite angular resolution causes source overlap on the sky, 
leading to confusion-induced blending that can bias photometry and reduce source detection completeness.
In addition, extended PSF wings and ghost\footnote{In optical terminology, 
stray light is broadly categorized into {\lqq}ghosts{\rqq} and {\lqq}flare{\rqq}. 
While flare generally refers to widely spread diffuse scattered light that does not form a distinct image, 
a ghost arises from discrete internal specular reflections 
and typically forms a relatively concentrated, sometimes nearly point-like structure. 
Because the stray light discussed in this work originates from multiple internal reflections 
and maintains a concentrated profile, we adopt the term {\lqq}ghost{\rqq}.}
halos redistribute flux from bright sources  to larger radii, 
and can raise the apparent local background level and increase source crowding, 
thereby exacerbating blending and increasing the background fluctuations.
Throughout this paper, we use the term source confusion to broadly include
both confusion noise and confusion-induced blending; 
their impacts on background fluctuations,
photometry, and source detection are collectively referred to as source confusion effects.
We define the confusion limit as the regime 
in which source confusion dominates the noise budget
and imposes an effective floor on the achievable limiting depth.
Once this regime is approached, further integration yields diminishing returns,
because the depth improvement is no longer driven primarily by instrumental or photon noise.
It is therefore important to quantify source confusion effects in advance
when designing a survey at long wavelengths
(\citealt{2001MNRAS.325.1241V}; \citealt{2004ApJS..154...93D}; \citealt{2006PASJ...58..673M}).

As a concrete example of incorporating the effects of source confusion into survey design,
\cite{2003ApJ...585..617D} have quantified confusion noise 
using a phenomenological model of galaxy evolution in advance of deep surveys 
with the Multiband Imaging Photometer for Spitzer (MIPS; \citealt{2004ApJS..154...25R}) 
on board the Spitzer Space Telescope (\citealt{2004ApJS..154....1W}).
They have estimated the expected sensitivities 
and showed that such surveys would be able to detect faint infrared sources (see also \citealt{2004ApJS..154...93D}).
In practice, \cite{2004ApJS..154...70P} and \cite{2006ApJ...647L...9F} 
have reported detections of faint sources with MIPS 
at $24\mu$m and $70\mu$m, respectively, 
and derived their number counts, 
successfully resolving more than half of the cosmic infrared background
(see also \citealt{2004ApJS..154...80C} and the review by \citealt{2008ARA&A..46..201S}).

With the advent of the James Webb Space Telescope
(JWST; \citealt{2006SSRv..123..485G}; \citealt{2023PASP..135f8001G}; \citealt{2023PASP..135e8001M}),
the study of distant galaxies has advanced dramatically over the past several years
(see the reviews by 
\citealt{2025NatAs...9.1134A}, 
\citealt{2025ConPh..66..116M}, 
and \citealt{2026enap....4..453S}; 
see also the recent lecture notes by \citealt{2025arXiv250816948E} 
for a broad overview of galaxy formation).
However, the JWST near-infrared camera NIRCam
(\citealt{2005SPIE.5904....1R}; \citealt{2023PASP..135b8001R})
has a relatively small field of view of $9.7$ arcmin$^2$, 
corresponding to about $2.7 \times 10^{-3}$ deg$^2$, 
which limits its efficiency for finding rare bright sources.
GREX-PLUS is one of the future missions expected to play a complementary role to JWST
by conducting wide-area surveys that leverage its much larger field of view.
At the same time, the smaller mirror of GREX-PLUS leads to a lower angular resolution than JWST,
making it more susceptible to source confusion effects.
Conveniently, JWST observes at wavelengths similar to GREX-PLUS.
In this paper, we therefore take an empirical approach to evaluating the impact of source confusion 
by constructing simulated GREX-PLUS images from deep high-resolution near-infrared images obtained with JWST.

This paper is organized as follows.
Section~\ref{sec:GREX-PLUS PSF and Ghost Model} presents 
the point spread function (PSF) and ghost patterns
computed from the latest GREX-PLUS optical design.
In Section~\ref{sec:Simulated Imaging Data for GREX-PLUS}, we construct simulated GREX-PLUS images
by convolving JWST images with the PSF$+$ghost kernel.
We then perform source extraction and evaluate the detection completeness
in Section~\ref{sec:analysis}.
Section~\ref{sec:results_and_discussion} compares the limiting magnitudes 
measured from the JWST and simulated GREX-PLUS images, 
provides a rough estimate of the GREX-PLUS integration times required to reach representative survey depths, 
derives completeness-corrected number counts,
and discusses the impact of source confusion.
Finally, Section~\ref{sec:summary} summarizes our results.
Throughout this paper, 
we use magnitudes in the AB system (\citealt{1983ApJ...266..713O}),
defined as $\mathrm{ABmag}=-2.5\log f_\nu - 48.60$,
where $f_\nu$ is the flux density in units of erg cm$^{-2}$ s$^{-1}$ Hz$^{-1}$.

\section{GREX-PLUS PSF and Ghost Model}
\label{sec:GREX-PLUS PSF and Ghost Model}

\hspace{0.8em}
The GREX-PLUS WFC comprises seven mercury cadmium telluride (HgCdTe) detector arrays,  
covering a wavelength range from $2\mu$m to $8\mu$m.
The planned broadband filters and their wavelength ranges are
F232 ($2.0$--$2.6\,\mu\mathrm{m}$),
F303 ($2.6$--$3.4\,\mu\mathrm{m}$),
F397 ($3.4$--$4.5\,\mu\mathrm{m}$),
F520 ($4.5$--$5.9\,\mu\mathrm{m}$), and
F680 ($5.9$--$7.7\,\mu\mathrm{m}$).
The WFC does not employ a filter wheel; 
each detector instead has a fixed bandpass filter.
These filters are made of zinc sulfide (ZnS) with a thickness of $5$~mm 
and are positioned at a distance of $2$~mm from the detector entrance surface.
Three detectors will be devoted to the F397 band,
while the remaining four detectors will be assigned one each to the other bands.
We allocate three detectors to F397 
to compensate for its reduced sensitivity 
since the zodiacal thermal background becomes significant at these wavelengths 
(\citealt{2023arXiv230408104G}).
In addition, the typical spectral energy distributions of high-$z$ galaxies 
are expected to be roughly flat in $f_{\nu}$ or decline toward longer wavelengths. 
Therefore, achieving a comparable depth at F397 to that in the shorter wavelength bands 
generally requires a longer effective exposure.
At longer wavelengths of F520 and F680, 
the primary role is instead to reject low-$z$ interlopers,
so a shallower depth is acceptable 
and a single detector is assigned to each band. 
Because source confusion effects generally become more severe toward longer wavelengths,
we focus in this work on the PSF and ghost behavior in the F397 band near $4\mu$m, 
which spans the longest wavelengths among the deep imaging bands.

\begin{figure}
\begin{center}
   \includegraphics[width=0.47\textwidth]{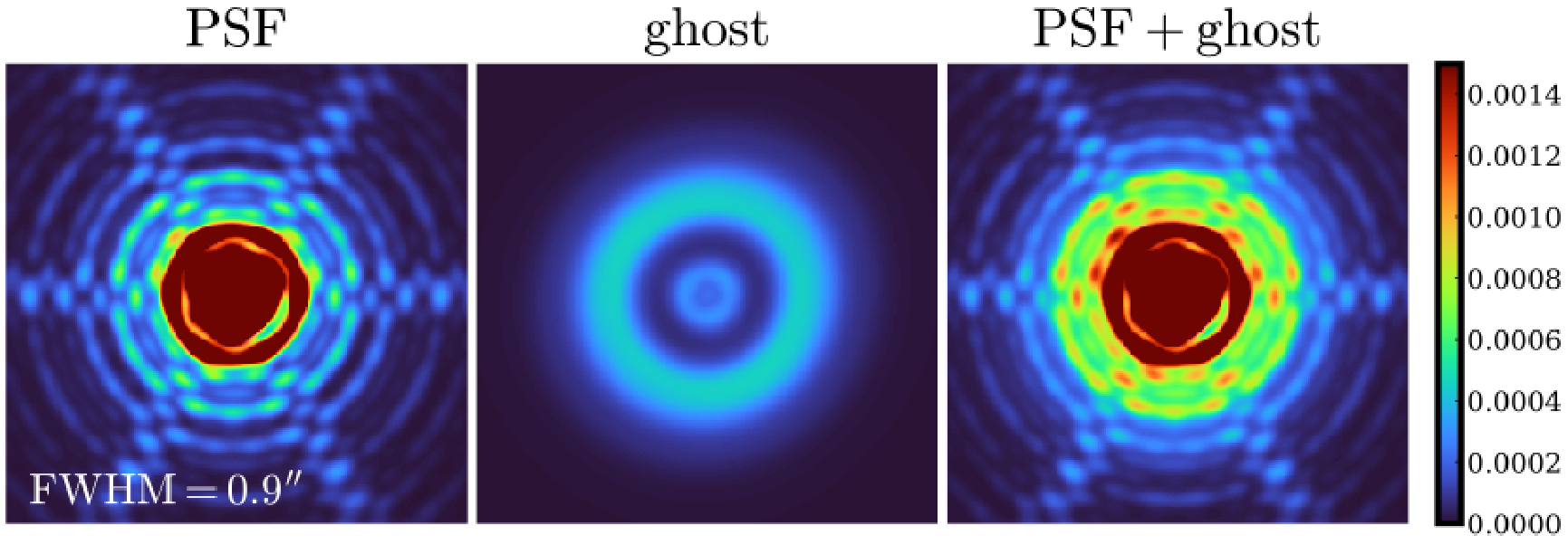}
   \includegraphics[width=0.47\textwidth]{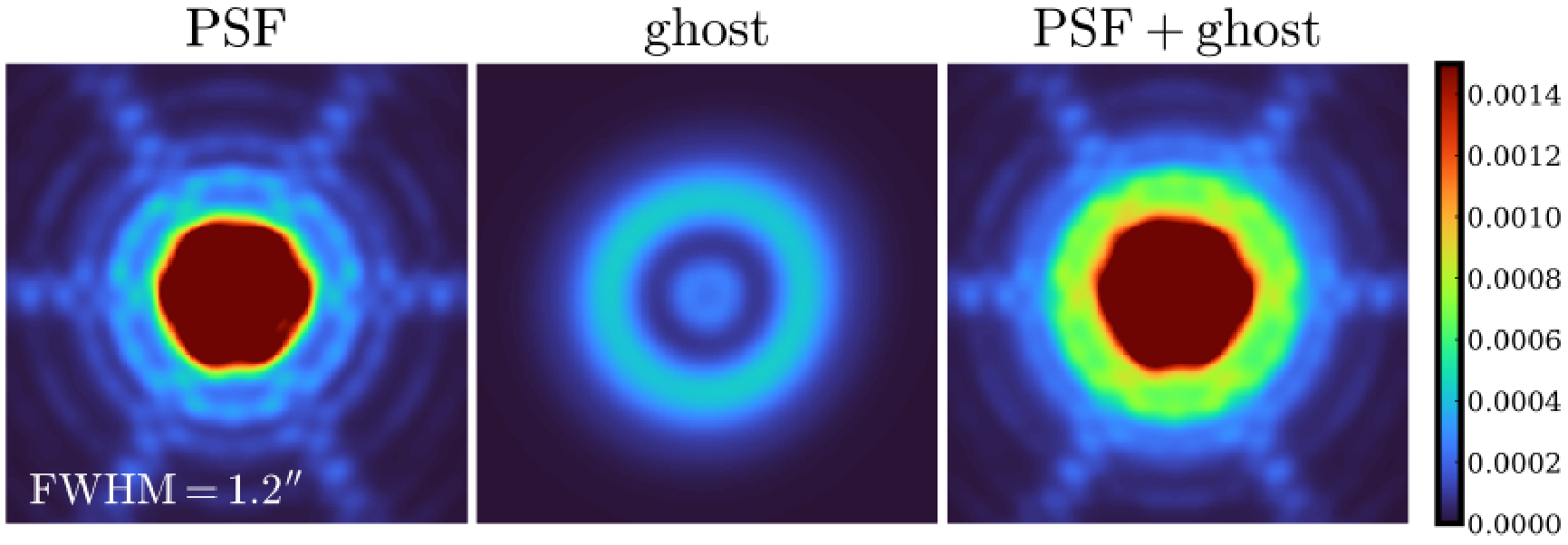}
\caption{
Images of the PSF and ghost calculated based on the GREX-PLUS optical model.
The top row shows the case with PSF FWHM = $0\farcs9$, 
where only pointing uncertainty is included, 
while the bottom row are the case with PSF FWHM = $1\farcs2$, 
where both pointing and optical uncertainties are included.
In each row, the panels from left to right denote  
the PSF image, the ghost image, and the combined PSF$+$ghost image.
All images span $20'' \times 20''$.
}
\label{fig:PSF_ghost_images}
\end{center}
\end{figure}

\begin{figure}
\begin{center}
   \includegraphics[width=0.37\textwidth]{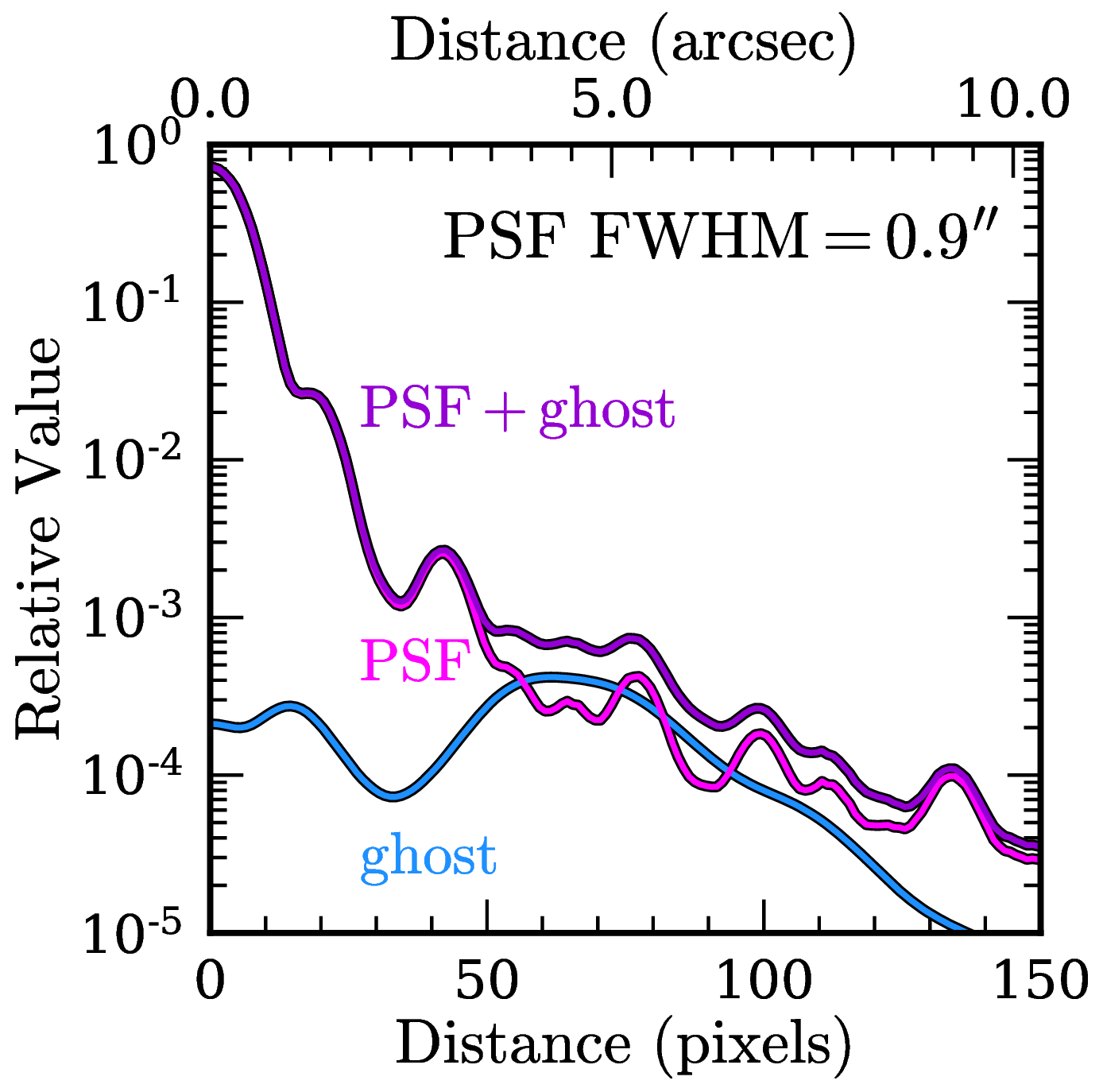}
   \includegraphics[width=0.37\textwidth]{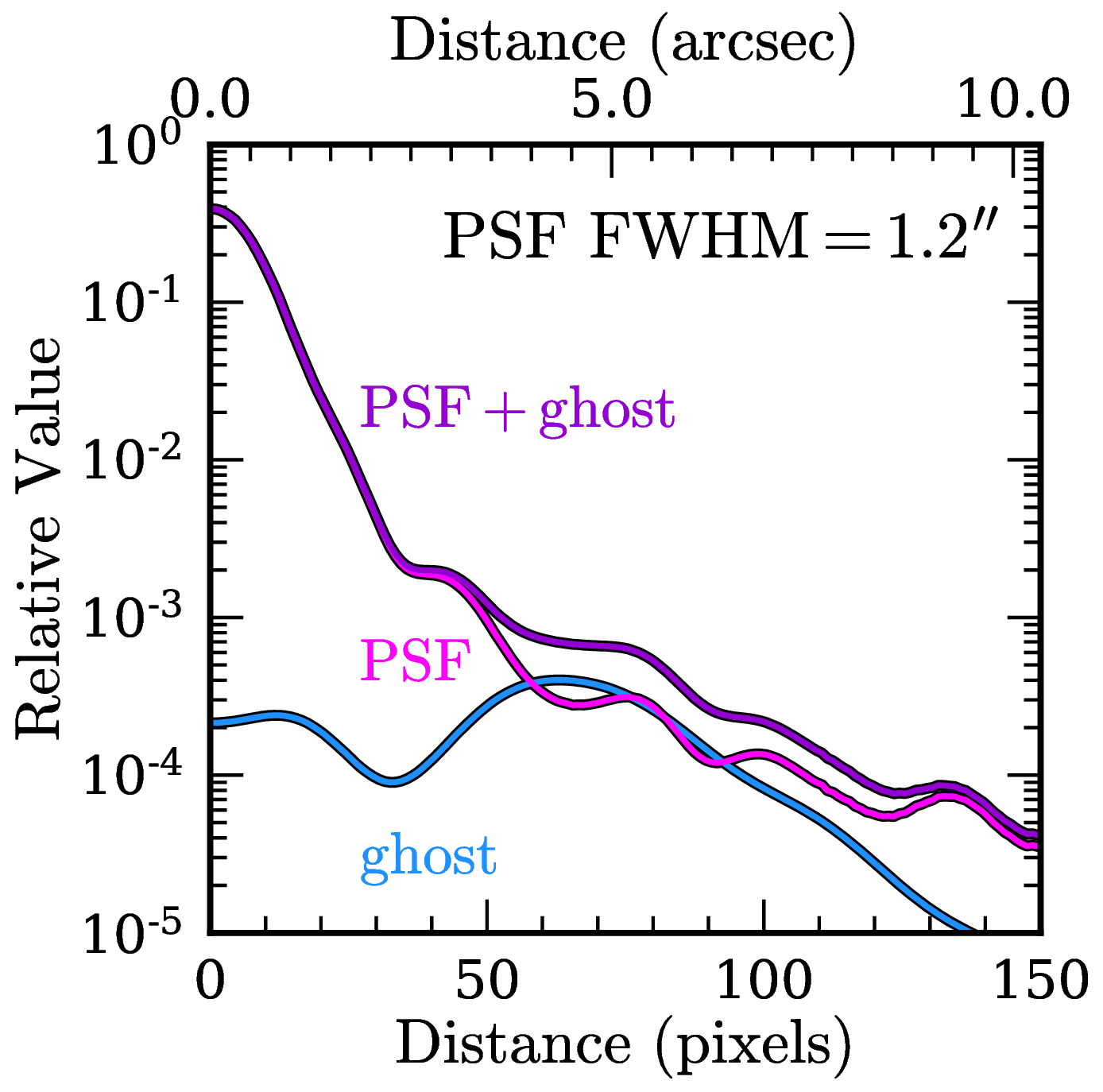}
\caption{
Radial profiles of the PSF and ghost calculated from Figure \ref{fig:PSF_ghost_images}.
The top panel presents the case with PSF FWHM = $0\farcs9$,  
and the bottom panel shows the case with PSF FWHM = $1\farcs2$.
The magenta curve denotes the PSF, 
the cyan curve indicates the ghost, 
and the purple curve is the combined PSF$+$ghost profile.
At radii around $5$ arcsec, 
the PSF and ghost contributions become comparable, 
whereas at other radii the PSF dominates.
}
\label{fig:PSF_radial_profile}
\end{center}
\end{figure}

Taking advantage of its wide field of view, 
GREX-PLUS is designed to carry out a wide-area imaging survey 
consisting of three layers: Deep, Medium, and Wide 
(\textcolor{blue}{GREX-PLUS Science Team et al. in preparation}). 
The Deep layer will uniformly map $10$~deg$^2$ 
with an effective integration time of $17.8$~hours per sky position, 
reaching a $5\sigma$ depth of $\simeq26.5$~mag at $4\mu$m. 
The Medium layer will uniformly map $100$~deg$^2$ 
with $2.8$~hours per sky position,
achieving $5\sigma \simeq 25.5$~mag at $4\mu$m. 
The Wide layer will uniformly map $1{,}000$~deg$^2$ 
with $600$~sec per sky position,
reaching $5\sigma \simeq24.0$~mag at $4\mu$m.

The PSF describes the response of the instrument to a point source, 
and its characteristics are essential for evaluating the effects of source confusion.
We compute the GREX-PLUS PSF from the latest optical design 
using a standard Fourier optics calculation.
For each representative field point, 
the PSF intensity is derived by Fourier-transforming the complex pupil function, 
which incorporates the wavefront-error map and the aperture geometry 
including obscurations from the secondary mirror and its spiders.

In practice, the final image quality is further broadened by pointing jitter 
and residual imperfections in the optical system (e.g., mirror figure and alignment tolerances).
For GREX-PLUS, the requirement on this additional broadening is 
$\sigma_{\rm jit} = 0.2$~arcsec from pointing jitter alone, 
and $\sigma_{\rm tot} = 0.4$~arcsec when both pointing jitter and the optical imperfections are considered.
If the achieved mirror surface quality, alignment accuracy, and other subsystem tolerances 
are better than the requirements, the actual broadening is expected to be smaller than $0.4$~arcsec.
We therefore adopt an effective Gaussian broadening in the range
$\sigma=0.2$--$0.4$~arcsec to be applied to the theoretical PSF.
Strictly speaking, such perturbations originate from wavefront errors 
and should be modeled at the complex amplitude level in the pupil plane. 
However, given that our objective is to phenomenologically bracket 
the expected range of image quality degradation, 
we adopt a Gaussian smoothing in the intensity domain as a practical approximation. 
Accordingly, in the following we examine two cases: 
(i) the optically computed PSF smoothed with a Gaussian of $\sigma = 0.2$ arcsec, 
and (ii) the same PSF smoothed with a Gaussian of $\sigma = 0.4$ arcsec.
The resulting PSF full width at half maximum (FWHM) values are 
approximately $0.9$ arcsec for the former case 
and $1.2$ arcsec for the latter.

The left panels of Figure \ref{fig:PSF_ghost_images} show the PSF images for the two cases, 
and Figure~\ref{fig:PSF_radial_profile} presents their radial profiles.
These PSF radial profiles exhibit a sequence of diffraction rings in the wings,   
originating from the circular telescope aperture. 
In the GREX-PLUS optical system, 
the central obscuration by the secondary mirror modifies this standard Airy-like pattern, 
notably enhancing the ring amplitudes at specific radii, 
most prominently at a distance of about $5.2$~arcsec from the center. 
Additionally, a six-spike diffraction pattern is produced by the three spider vanes 
supporting the secondary mirror; its strength depends on the vane width,
which is set to $20$~mm in the optical model.

\begin{figure*}
\centering
\resizebox{\textwidth}{!}{%
{\sffamily
\begin{tikzpicture}[
  line cap=round, line join=round,
  plane/.style={line width=1.0pt},
  ray/.style={line width=1.0pt},
  arr/.style={-{Stealth[length=3.2mm,width=2.6mm]}, line width=1.0pt},
  title/.style={font=\large}
]
\def\L{2.0}     
\def\yA{0.0}    
\def\yB{-2.0}   
\def\yC{-3.0}   
\def\yText{-4.3} %
\def\dx{6.0}    

\begin{scope}[shift={(0,0)}]
  \node[title] at (0,1.1) {Nominal path};

  \draw[plane] (-\L,\yA) -- (\L,\yA);
  \draw[plane] (-\L,\yB) -- (\L,\yB);
  \draw[plane] (-\L,\yC) -- (\L,\yC);

  \node[anchor=east] at (-\L-0.25,\yA) {Filter front};
  \node[anchor=east] at (-\L-0.25,\yB) {Filter back};
  \node[anchor=east] at (-\L-0.25,\yC) {Detector entrance};

  \draw[arr] (0,0.6) -- (0,-3.8);

  \node[anchor=east] at (-\L+1.4,\yText) {Throughput:};

  \node[font=\large] at (0,\yText) {$t_1 \, t_2 \, t_3$}; 
\end{scope}

\begin{scope}[shift={(\dx,0)}]
  \node[title] at (0,1.1) {Path 1};

  \draw[plane] (-\L,\yA) -- (\L,\yA);
  \draw[plane] (-\L,\yB) -- (\L,\yB);
  \draw[plane] (-\L,\yC) -- (\L,\yC);

  \draw[arr] (-0.8,\yA+0.6) -- (-0.8,\yC); 
  \draw[arr] (-0.1,\yC) -- (-0.1,\yB); 
  \draw[arr] (0.6,\yB) -- (0.6,-3.8); 

  \node[font=\large] at (0,\yText) {$t_1 \, t_2 \, r_3 \, r_2 \, t_3$}; 
\end{scope}

\begin{scope}[shift={(2*\dx,0)}]
  \node[title] at (0,1.1) {Path 2};

  \draw[plane] (-\L,\yA) -- (\L,\yA);
  \draw[plane] (-\L,\yB) -- (\L,\yB);
  \draw[plane] (-\L,\yC) -- (\L,\yC);

  \draw[arr] (-0.8,\yA+0.6) -- (-0.8,\yB); 
  \draw[arr] (-0.1,\yB) -- (-0.1,\yA); 

  \draw[arr] (0.6,\yA) -- (0.6,-3.8);

  \node[font=\large] at (0,\yText) {$t_1 \, r_2 \, r_1 \, t_2 \, t_3$}; 
\end{scope}

\begin{scope}[shift={(3*\dx,0)}]
  \node[title] at (0,1.1) {Path 3};

  \draw[plane] (-\L,\yA) -- (\L,\yA);
  \draw[plane] (-\L,\yB) -- (\L,\yB);
  \draw[plane] (-\L,\yC) -- (\L,\yC);

  \draw[arr] (-0.8,\yA+0.6) -- (-0.8,\yC); 
  \draw[arr] (-0.1,\yC) -- (-0.1,\yA); 
  \draw[arr] (0.6,\yA) -- (0.6,-3.8); 
  \node[font=\large] at (0,\yText) {$t_1 \, t_2 \, r_3 \, t_2 \, r_1 \, t_2 t_3$}; 
\end{scope}

\end{tikzpicture}
}
}
\caption{Schematic light paths for the nominal light path 
transmitted through the filter and detector (Nominal path),
and the three dominant ghost-producing paths 
caused by reflections at the front and back surfaces of the filter and at the detector surface 
(Path 1 -- Path 3).
The relevant surfaces are labeled as
$i=1$ (Filter front), $i=2$ (Filter back), and $i=3$ (Detector entrance surface),
with transmission and reflection coefficients $t_i$ and $r_i$, respectively. 
Their adopted numerical values are listed in Table~\ref{tab:ghost_coeff}.
The paths shown here are the dominant terms retained up to second order in the reflection coefficients.
For each path, the calculated throughput is presented at the bottom.
}
\label{fig:ghost_reflections}
\end{figure*}
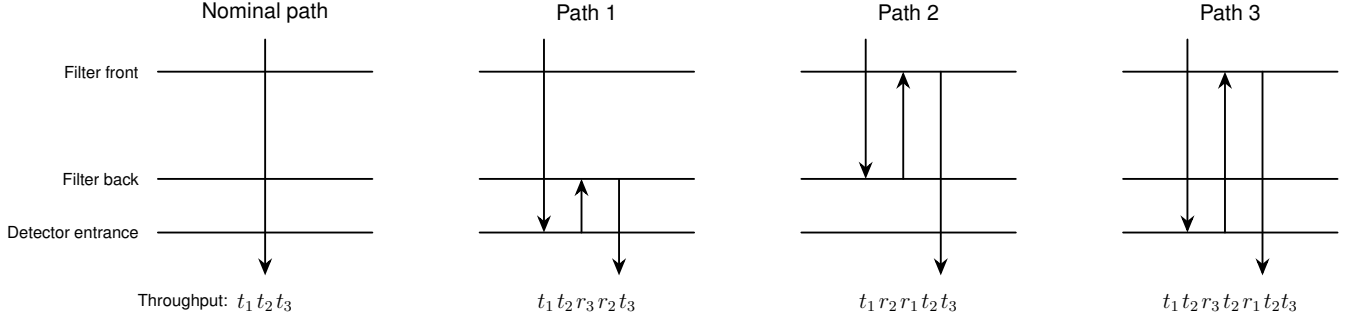

\begin{table}
\caption{Adopted Transmission Coefficient $t_i$ and Reflection Coefficient $r_i$ 
Used to Estimate Ghost Throughputs}
\begin{center}
{\footnotesize
\begin{tabular}{lcc} \hline
Surface $i$ & $t_i$ & $r_i$ \\ \hline
1: Filter front surface & $0.9$ & $0.1$ \\
2: Filter back surface  & $0.9$ & $0.1$ \\
3: Detector entrance surface & $0.8$ & $0.2$ \\ \hline
\end{tabular}
}
\end{center}
{\footnotesize Note.
For the detector entrance surface, we adopt an effective reflectivity of $r_3=0.2$.
The remaining fraction, $t_3=0.8$, is assumed to be absorbed in the detector and thus contributes to the nominal signal.
}
\label{tab:ghost_coeff}
\end{table}

In addition to the nominal PSF, 
internal reflections in the instrument (e.g., at the detector and filter surfaces) produce ghost components,
which are also important for the assessment of source confusion effects.
We compute the two-dimensional ghost profile based on the latest GREX-PLUS optical model.
For the F397 band, we adopt a band-averaged ghost kernel, obtained by averaging the model ghost images
computed from $\lambda=3.4\,\mu$m to $4.5\,\mu$m in steps of $0.1\,\mu$m, corresponding to the filter bandpass.
To use the PSF and ghost components consistently in our convolution kernel,
we estimate the relative brightness of the ghost with respect to the PSF
by comparing the ghost-producing reflection paths with the nominal optical path
that produces the PSF as outlined below.

The left panel of Figure~\ref{fig:ghost_reflections} illustrates the nominal optical path,
in which light is transmitted through the front and back surfaces of the filter and reaches the detector,
while the remaining panels show three representative reflection paths that produce ghosts.
We estimate the relative ghost throughputs by treating each surface interaction
with transmission and reflection coefficients.
Because the additional optical path length introduced by these multiple reflections
is sufficiently longer than the coherence length determined by the bandpass of the filter,
interference effects can be safely neglected.
Therefore, the ghost contributions can be accurately evaluated through the incoherent addition of light intensities.
To make the notation compact, we label the relevant surfaces as 
$i=1$ (filter front), $i=2$ (filter back), and $i=3$ (detector entrance surface),
and denote their transmission and reflection coefficients as $t_i$ and $r_i$, respectively, 
with the adopted values summarized in Table~\ref{tab:ghost_coeff}.
For the nominal path, the throughput is
\begin{equation}
T_{\rm n} = t_1\, t_2\, t_3 .
\end{equation}
In principle, multiple higher-order reflections generate an infinite series of ghost paths.
Here we consider $r_i \ll t_i$ and retain only the leading contributions up to second order in reflectivity,
i.e., paths involving two reflections.
For Path~1 (detector reflection followed by a reflection at the back filter surface), the throughput is
\begin{equation}
T_{1} = t_1\, t_2\, r_3\, r_2\, t_3 ,
\end{equation}
so that $T_{1}/T_{\rm n} = r_2 r_3 \simeq 2.0$\,\%.
For Path~2 (two reflections between the filter surfaces), the throughput is
\begin{equation}
T_{2} = t_1\, r_2\, r_1\, t_2\, t_3 ,
\end{equation}
giving $T_{2}/T_{\rm n} = r_1 r_2 \simeq 1.0$\,\%.
For Path~3 (detector reflection followed by a reflection at the front filter surface),
the light traverses the back filter surface twice on the round trip,
and the throughput is approximated as 
\begin{equation}
T_{3} = t_1\, t_2\, r_3\, (t_2\, r_1\, t_2)\, t_3
      = t_1\, t_2^{3}\, r_3\, r_1\, t_3 ,
\end{equation}
yielding $T_{3}/T_{\rm n} = t_2^{2} r_1 r_3 \simeq 1.6$\,\%.

Strictly speaking, each reflection path can produce a distinct ghost morphology. 
For example, the detailed calculation of the ghost halo based on the optical model indicates that,
compared with Path~1 and Path~2, the ghost halo produced by Path~3 has a radius 
that is larger by approximately a factor of two and is therefore more diffuse.
Paths involving more reflections have even lower throughputs, 
suppressed by additional factors of the reflectivities, and are thus expected to be sub-dominant.
We therefore adopt a fiducial estimate 
in which the total ghost flux is set to $2.0 + 1.0 + 1.6 = 4.6$\,\% of the PSF flux in the nominal path.
For simplicity, we represent the ghost using the Path~1 two-dimensional morphology,
which corresponds to the brightest individual reflection path, 
scaled so that its total flux is $4.6$\,\% of the nominal PSF flux.

\begin{table*}
\caption{$5\sigma$ Limiting Magnitudes of the JWST NIRCam Images and the Simulated GREX-PLUS Images}
\begin{center}
\begin{tabular}{cccc} \hline
Field				& JWST	& \multicolumn{2}{c}{simulated GREX-PLUS} \\ 
				& 		& (PSF FWHM = $0\farcs9$)	& (PSF FWHM = $1\farcs2$) \\\hline
JADES GOODS-S	& $30.86$ / $28.53$ / $28.21$	& $27.57$				& $27.07$	\\
JADES GOODS-N	& $30.20$ / $27.49$ / $27.19$	& $27.32$				& $26.89$	\\
GLASS			& $29.91$ / $26.76$ / $26.55$	& $26.75$				& $26.26$	\\
CEERS			& $29.77$ / $26.82$ / $26.64$	& $27.06$				& $26.75$	\\
COSMOS-Web		& $28.91$ / $25.95$ / $25.69$	& $26.19$				& $25.83$	\\
\hline
\end{tabular}
\end{center}
{\footnotesize Note. 
For the JWST images, 
the three values from left to right correspond to the limiting magnitudes 
measured using random circular apertures 
with diameters of $0\farcs2$, $2\farcs0$, and $2\farcs4$, respectively.
For the simulated GREX-PLUS images, 
we adopt apertures with diameters approximately twice the PSF FWHM: 
$2\farcs0$ for the case with PSF FWHM $=0\farcs9$, 
and $2\farcs4$ for the case with PSF FWHM $=1\farcs2$.
}
\label{tab:limitmag}
\end{table*}

The middle panels of Figure~\ref{fig:PSF_ghost_images} show the two-dimensional ghost images 
calculated from the GREX-PLUS optical model, 
scaled to the adopted relative flux with respect to the nominal PSF.
For consistency with the PSF treatment, 
we also smooth the ghost images with the same Gaussian broadenings,
$\sigma=0.2$ and $0.4$~arcsec, in the two cases considered here.
Figure~\ref{fig:PSF_radial_profile} presents the corresponding radial profiles of the ghosts.
As seen from the radial profiles, the PSF contribution dominates over the ghost at almost all radii.
At radii of about $5.0$ arcsec, the ghost and PSF contributions become comparable.

In the right panels of Figure \ref{fig:PSF_ghost_images},
we present the combined images of the PSF and ghost 
for the two cases with PSF FWHM values of $0.9$ arcsec and $1.2$ arcsec.
For simplicity, we construct the PSF$+$ghost convolution kernel 
by co-centering the ghost with the nominal PSF.
In practice, toward the edge of the field, 
the incident rays can have a finite angle relative to the surface normals of the filter and detector, 
so that the ghost image is expected to be displaced from the nominal PSF position. 
We neglect this positional offset in the present analysis.
Figure \ref{fig:PSF_radial_profile} also show their radial profiles together.
In the next section, we use the PSF$+$ghost images 
as the convolution kernels to generate simulated GREX-PLUS images.
We adopt a kernel size of $20'' \times 20''$, 
which fully covers the region where the PSF and ghost contributions become comparable.
Although the PSF and ghost formally extend to larger radii, 
a much larger kernel would substantially increase the computational cost of the convolutions.
From a larger PSF$+$ghost image spanning about $10$ arcmin on a side,
we estimate that the flux outside this kernel is only $\simeq 3${\%} of the total. 
We therefore neglect this small contribution in the following analysis.

\section{Simulated Imaging Data for GREX-PLUS}
\label{sec:Simulated Imaging Data for GREX-PLUS}

\hspace{0.8em}
To generate simulated images expected for GREX-PLUS,
we take an empirical approach.
We use real JWST NIRCam imaging data 
that cover similar wavelengths to GREX-PLUS but with much higher spatial resolution.
By convolving NIRCam images from surveys spanning a range of depths
with the GREX-PLUS PSF$+$ghost kernel,
we construct simulated GREX-PLUS images corresponding to different effective depths.
We then measure the limiting magnitudes of these simulated images
to investigate how the limiting magnitude improves with integration time
down to sensitivities comparable to or deeper than the planned GREX-PLUS survey. 
Using this approach, we also quantify the impact of source confusion
by examining whether the simulated GREX-PLUS images
continue to deepen as the input JWST images become deeper.

Specifically, we use NIRCam images taken in the following four surveys:
JWST Advanced Deep Extragalactic Survey
(JADES; 
\citealt{2023ApJS..269...16R}; \citealt{2024A&A...690A.288B}; \citealt{2024ApJ...964...71H}; 
\citealt{2025ApJS..277....4D}; \citealt{2025ApJS..281...50E}; \citealt{2026ApJS..283....6E}),
JWST Grism Lens-Amplified Survey from Space
(GLASS; \citealt{2022ApJ...935..110T}; \citealt{2023ApJ...952...20P};
\citealt{2024A&A...690A...2M}; \citealt{2025A&A...699A.225W}),
Cosmic Evolution Early Release Science survey
(CEERS; \citealt{2025ApJ...983L...4F};
see also \citealt{2022ApJ...940L..55F}; \citealt{2023ApJ...946L..13F};
\citealt{2023ApJ...946L..12B}),
and COSMOS-Web (\citealt{2023ApJ...954...31C}; \citealt{2026ApJ...999..200F}).
For JADES, 
we use the official JADES NIRCam images available on MAST,\footnote{\url{https://archive.stsci.edu/hlsp/jades}}
focusing on an especially deep $\simeq 2.3$~arcmin$^2$ region in GOODS-S 
and a $\simeq 3.4$~arcmin$^2$ region in GOODS-N. 
For GLASS and CEERS, 
we use NIRCam images reduced by \cite{2023ApJS..265....5H}.  
The GLASS field covers $\simeq 8.4$ arcmin$^2$.
For CEERS, we use the CEERS2 region, covering $\simeq 9.2$ arcmin$^2$.
For COSMOS-Web, 
we download NIRCam data for the A2 tile covering $\simeq 25$ arcmin$^2$ 
from their official webpage.\footnote{\url{https://cosmos2025.iap.fr/}} 
For JADES, GLASS, and CEERS, we use the NIRCam F356W images, 
where F356W is a broadband filter centered near $3.6\mu$m, 
i.e., close to $4\mu$m.
Since COSMOS-Web does not include F356W imaging,
we instead use the F277W image, which is nearby in wavelength
and provides slightly higher angular resolution.

For each JWST image, we estimate the $5\sigma$ limiting magnitudes
by placing random circular apertures in blank regions avoiding detected objects
and performing aperture photometry.
We adopt three different aperture diameters: $0\farcs2$, $2\farcs0$, and $2\farcs4$.
The $0\farcs2$ aperture is used to verify consistency with previous studies,
while the $2\farcs0$ and $2\farcs4$ apertures are chosen for direct comparisons 
with the results from the simulated GREX-PLUS images later in our analysis.
We construct the distribution of the measured aperture fluxes 
and use its standard deviation to estimate the background noise level, 
from which we derive the $5\sigma$ limiting magnitudes.
The derived $5\sigma$ limiting magnitudes are summarized in Table~\ref{tab:limitmag}.
For reference, the total on-source integration times of the original JWST images 
used here span more than an order of magnitude.
For JADES, the GOODS-N image has an integration time of 
$5.67\times10^3$~s ($\simeq 1.6$~hr; Table~3 of \citealt{2026ApJS..283....6E}),
while the GOODS-S image reaches $51.7\times10^3$~s 
($\simeq 14.4$~hr; Table~4 of \citealt{2026ApJS..283....6E}).
For GLASS, the effective integration time per sky position is 
$6$,$120$~s ($1.7$~hr; Table~1 of \citealt{2023ApJ...952...20P};
see also \citealt{2023ApJS..265....5H}). 
For CEERS, the corresponding value is $3$,$092$~s 
($\simeq 0.9$~hr; Table~2 of \citealt{2025ApJ...983L...4F}).
For COSMOS-Web, it is $257 \times 4 = 1$,$028$~s
($\simeq 0.3$~hr; Section~2.1 of \citealt{2026ApJ...999..200F}).

\begin{figure*}
\begin{center}
   \includegraphics[width=1.0\textwidth]{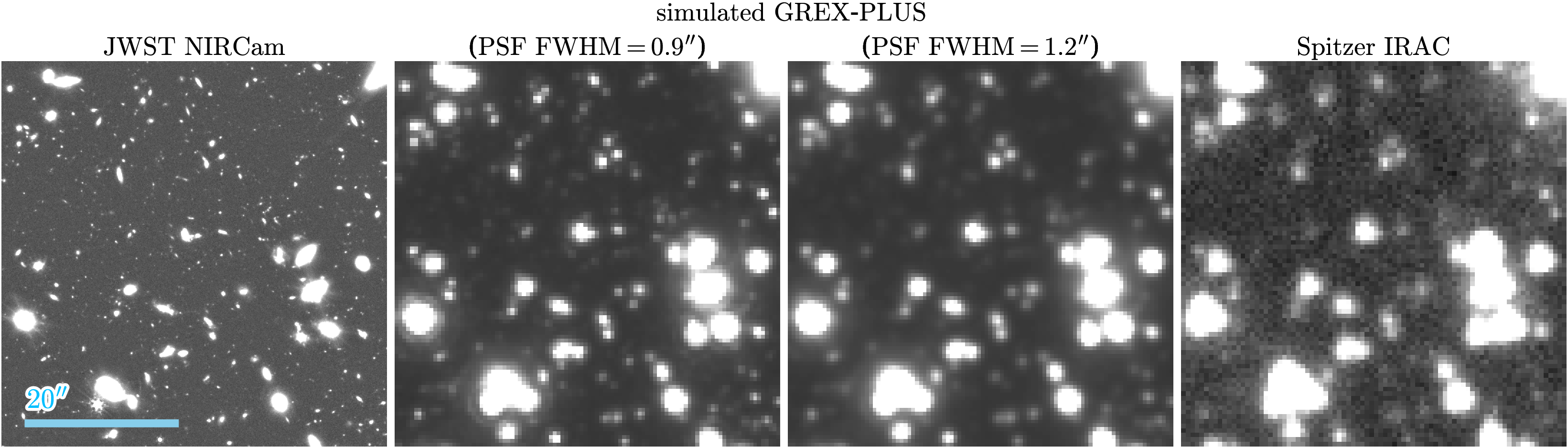}
\caption{Example simulated GREX-PLUS images created from a subregion of the JADES GOODS-S field.
From left to right, the panels show the original JWST NIRCam F356W image,
the simulated GREX-PLUS image generated by convolving the NIRCam image
with the PSF$+$ghost kernel for the case with PSF FWHM $=0\farcs9$,
the simulated GREX-PLUS image generated in the same manner for the case with PSF FWHM $=1\farcs2$, 
and, for comparison, the Spitzer IRAC $3.6\,\mu$m image. 
The horizontal bar in the lower left corner indicates a scale of $20''$.
The $5\sigma$ limiting magnitudes are $28.53$ and $28.21$~mag for the JWST image
measured in circular apertures with diameters of $2\farcs0$ and $2\farcs4$, respectively;
$27.57$~mag for the simulated GREX-PLUS image with PSF FWHM $=0\farcs9$
measured with $2\farcs0$-diameter apertures;
$27.07$~mag for the simulated GREX-PLUS image with PSF FWHM $=1\farcs2$
measured with $2\farcs4$-diameter apertures;
and $24.86$~mag for the Spitzer IRAC image measured with $3\farcs0$-diameter apertures.
}
\label{fig:simulated_GREX_PLUS_image}
\end{center}
\end{figure*}

The pixel scales of the JWST NIRCam images are either $0.015$ or $0.030$ arcsec pix$^{-1}$.
We first resample the NIRCam images to match 
the pixel scale of the GREX-PLUS PSF$+$ghost kernel,
$0.06895$ arcsec pix$^{-1}$, using the IRAF task \texttt{magnify}.
We then convolve the resampled images with the GREX-PLUS PSF$+$ghost kernel
using \texttt{convolve2d} in \texttt{scipy.signal}.
We generate simulated images for both cases 
with PSF FWHM values of $0.9$ arcsec and $1.2$ arcsec.
The original JWST NIRCam PSF has an FWHM of slightly above $0.1$ arcsec,
which is much smaller than the GREX-PLUS PSF.
Therefore, for simplicity, we skip deconvolution of the original NIRCam PSF and convolve the images directly.
Finally, we resample the convolved images to the nominal GREX-PLUS pixel scale,
$0.48$ arcsec pix$^{-1}$, again using the IRAF task \texttt{magnify}.
Note that this procedure does not add any additional noise terms specific to GREX-PLUS
(e.g., photon noise and detector noise), 
but simply redistributes the signal and the noise already present in the JWST images 
through resampling and convolution.

For each field, we estimate the $5\sigma$ limiting magnitudes 
of the simulated GREX-PLUS images 
by performing random aperture photometry in blank regions avoiding detected sources.
We adopt circular apertures with diameters approximately twice the PSF FWHM at random positions. 
Specifically, we use apertures with a diameter of $2\farcs0$ 
for the case with PSF FWHM $=0.9$~arcsec 
and $2\farcs4$ for the case with PSF FWHM $=1.2$~arcsec.
For reference, when the PSF$+$ghost kernel is measured within these apertures,
the enclosed flux fractions relative to the total flux are $70$\,\% and $71$\,\%, respectively,
corresponding to aperture corrections of $2.5\log(0.7) \simeq -0.4$~mag,  
which can be applied when converting point source aperture magnitudes to total magnitudes.
We then measure the background fluctuations.
The derived $5\sigma$ limiting magnitudes are listed in Table~\ref{tab:limitmag}.
We compare these limits with those measured from the original JWST images
in Section~\ref{subsec:Limiting Magnitude Comparison}.

\begin{table}
\caption{Summary of the SExtractor Source Extraction Parameters 
Adopted for the JWST Images and the Simulated GREX-PLUS Images}
\begin{center}
{\footnotesize
\begin{tabular}{ccc} \hline
Parameter & JWST & Simulated GREX-PLUS \\ \hline
\texttt{DETECT\_MINAREA}   & $5$   & $3$ \\
\texttt{DETECT\_THRESH}    & $1.8$ & $1.0$ \\
\texttt{ANALYSIS\_THRESH}  & $1.8$ & $1.0$ \\
\texttt{DEBLEND\_NTHRESH}  & $32$  & $32$ \\
\texttt{DEBLEND\_MINCONT}  & $0.005$  & $0.0001$ \\ \hline
\end{tabular}
}
\end{center}
{\footnotesize Note.
As explained in the main text, we use relatively permissive detection thresholds for source extraction
to include faint candidates in the initial catalog.
We define detected sources as those with aperture magnitudes brighter than the $5\sigma$ limiting magnitude.
}
\label{tab:sextractor_params}
\end{table}

\begin{figure*}
\begin{center}
   \includegraphics[width=0.4\textwidth]{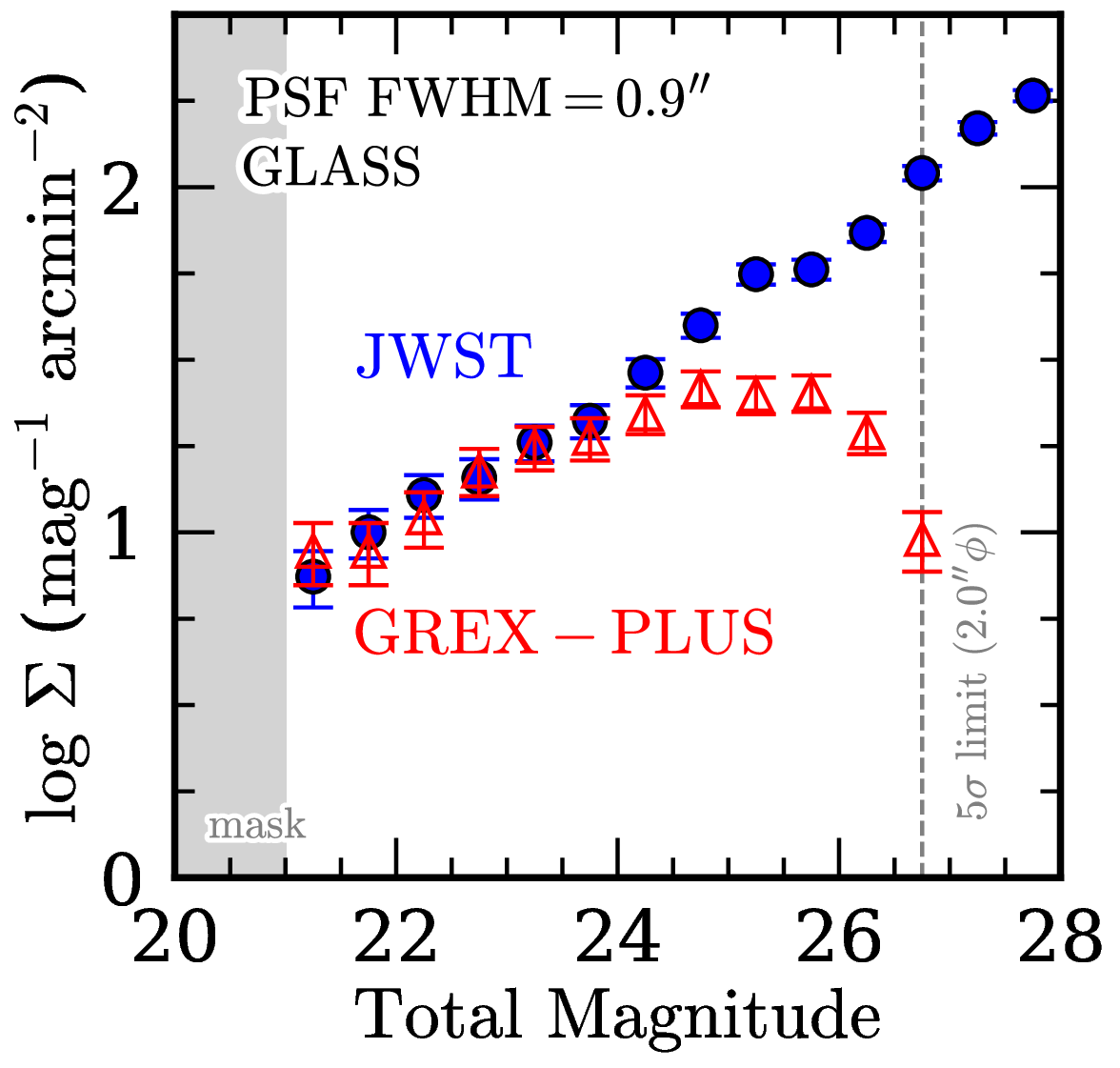}
   \includegraphics[width=0.4\textwidth]{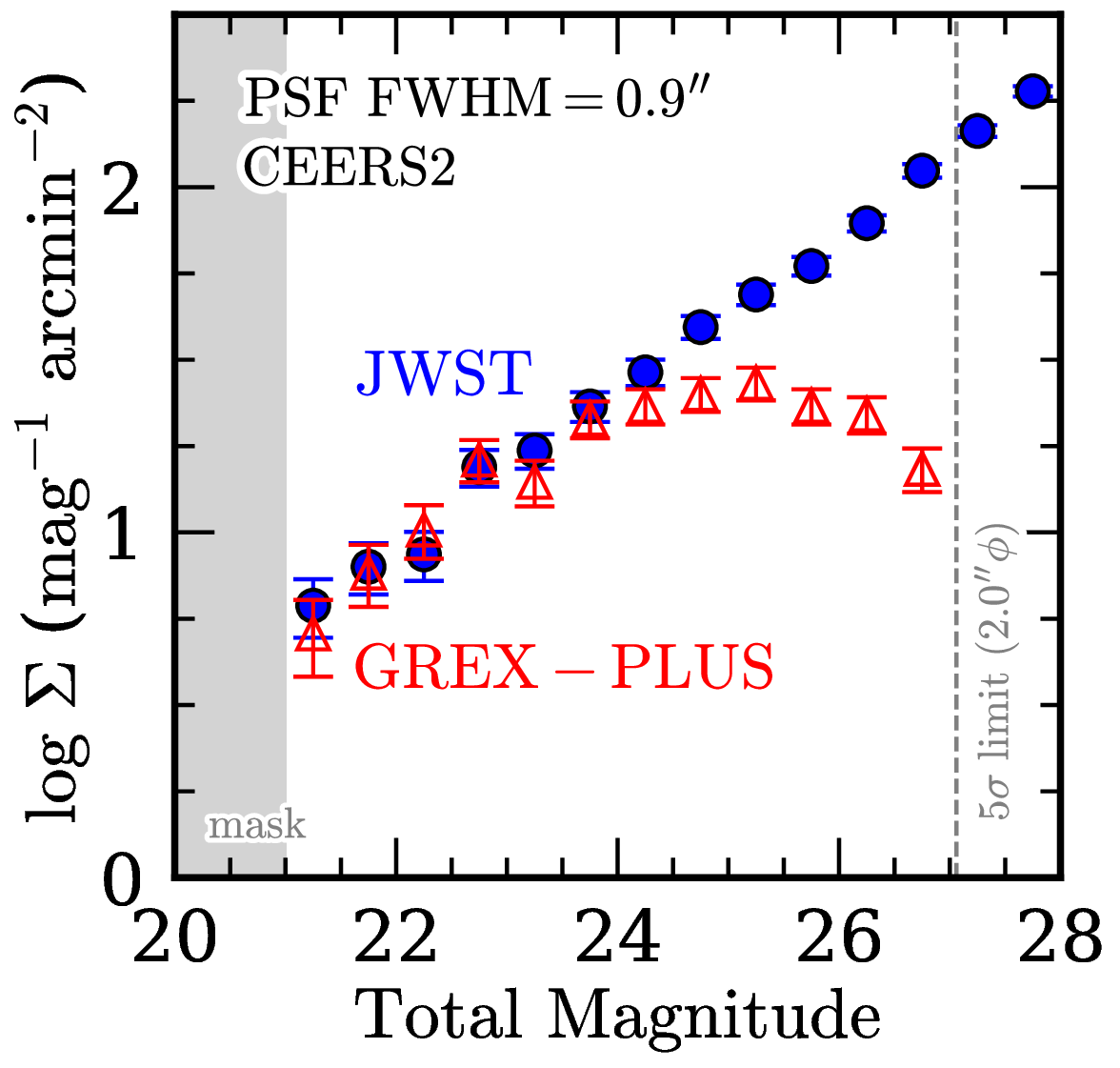}
   \includegraphics[width=0.4\textwidth]{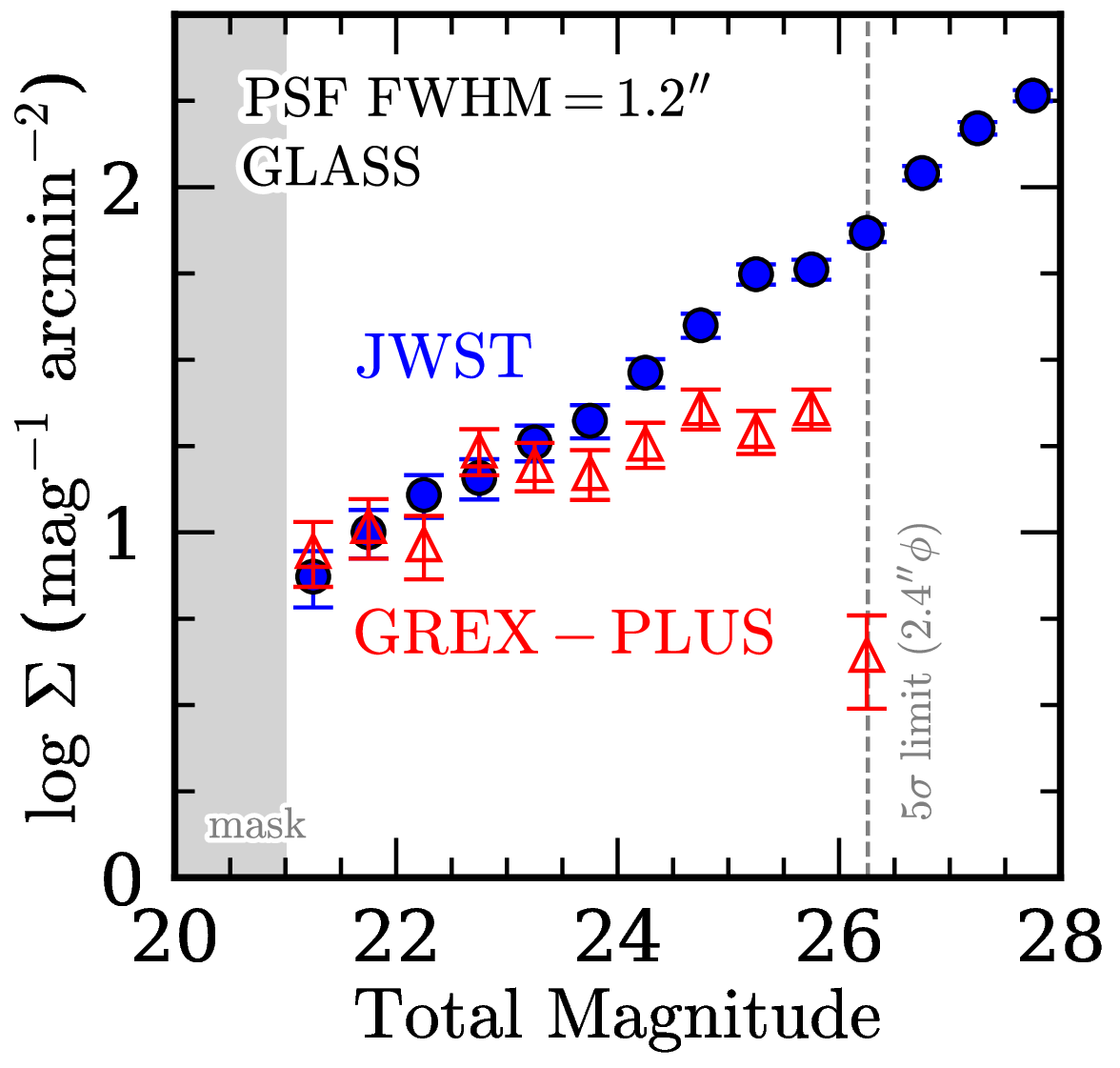}
   \includegraphics[width=0.4\textwidth]{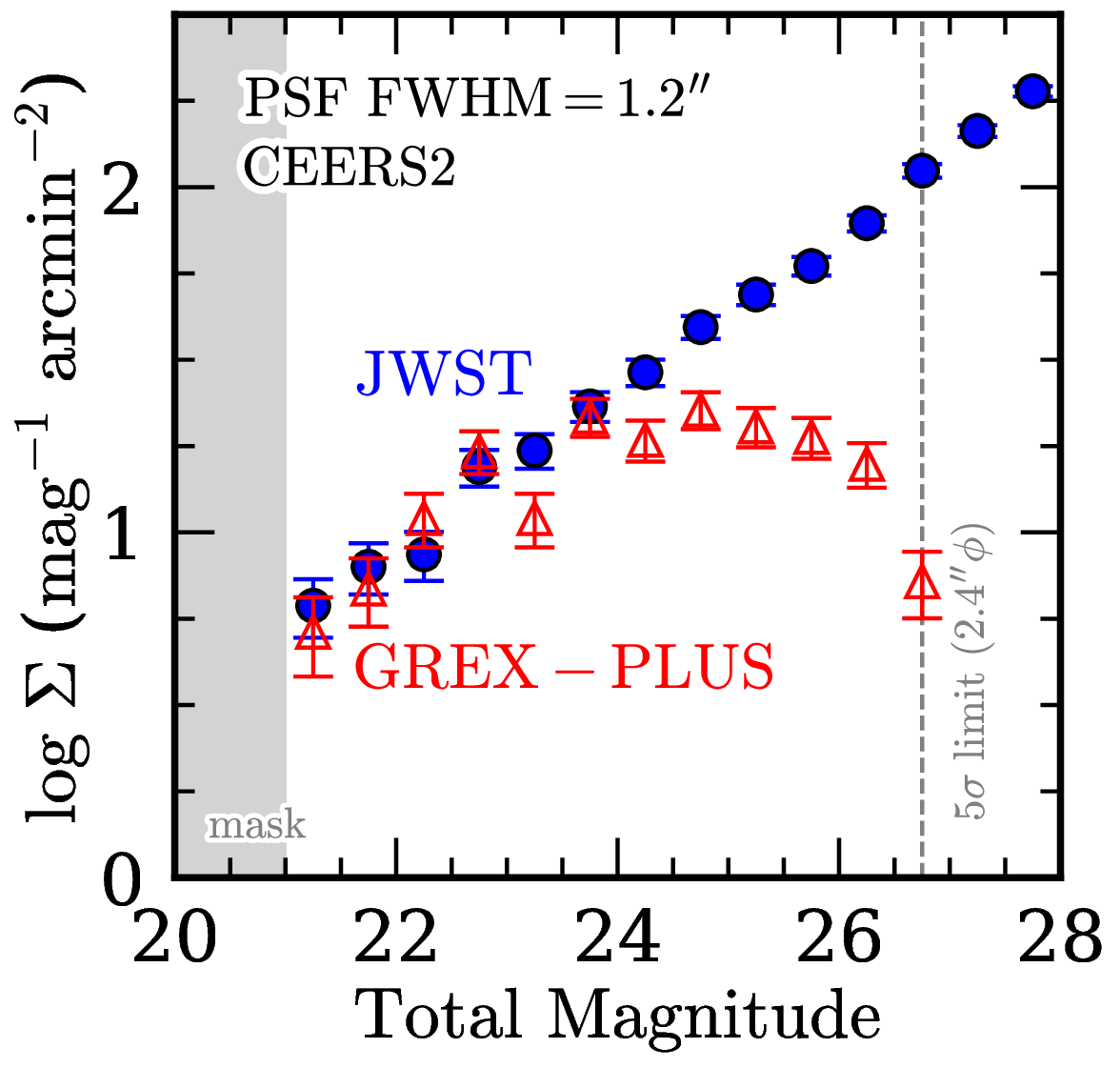}
\caption{
Number counts of sources with aperture magnitudes brighter than the $5\sigma$ limit,
detected in the simulated GREX-PLUS images and in the original JWST images.
These number counts are not corrected for detection incompleteness.
Red open triangles denote the results from the simulated GREX-PLUS images,
and blue circles are those from the JWST images.
The left panels show the GLASS field and the right panels are the CEERS field.
The top panels correspond to the case with PSF FWHM $=0\farcs9$,
and the bottom panels to PSF FWHM $=1\farcs2$.
Sources brighter than $21$~mag, including their surrounding PSF wings and ghost features,
are masked and are therefore not plotted.
The vertical gray dashed lines indicate the $5\sigma$ limiting aperture magnitude.
For the case with PSF FWHM $=0\farcs9$, 
the limiting magnitude is derived using circular apertures with a diameter of $2\farcs0$,
while for the case with PSF FWHM $=1\farcs2$, 
it is derived using circular apertures with a diameter of $2\farcs4$.
}
\label{fig:number_counts_GLASS}
\end{center}
\end{figure*}

\begin{figure*}
\begin{center}
   \includegraphics[width=0.4\textwidth]{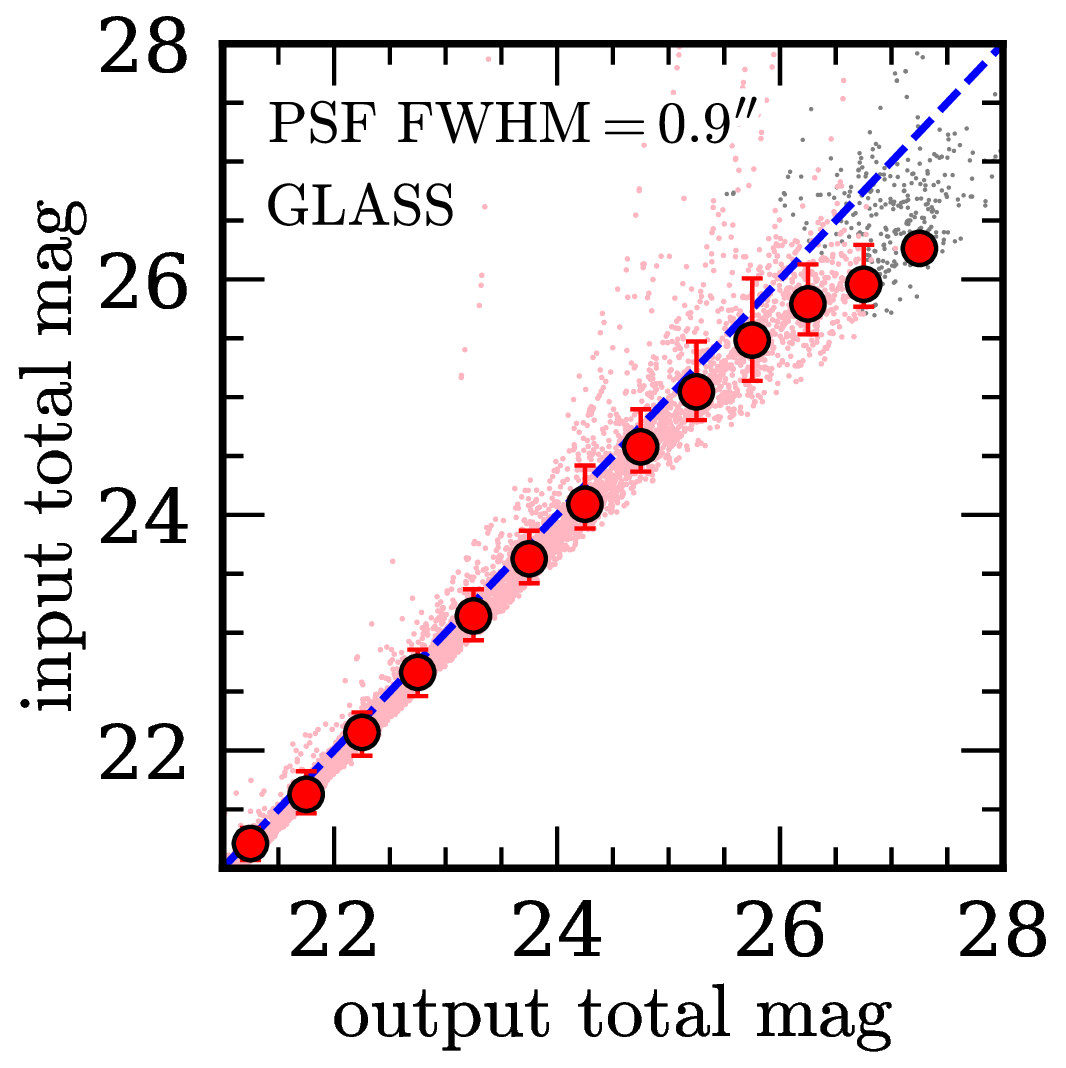}
   \includegraphics[width=0.4\textwidth]{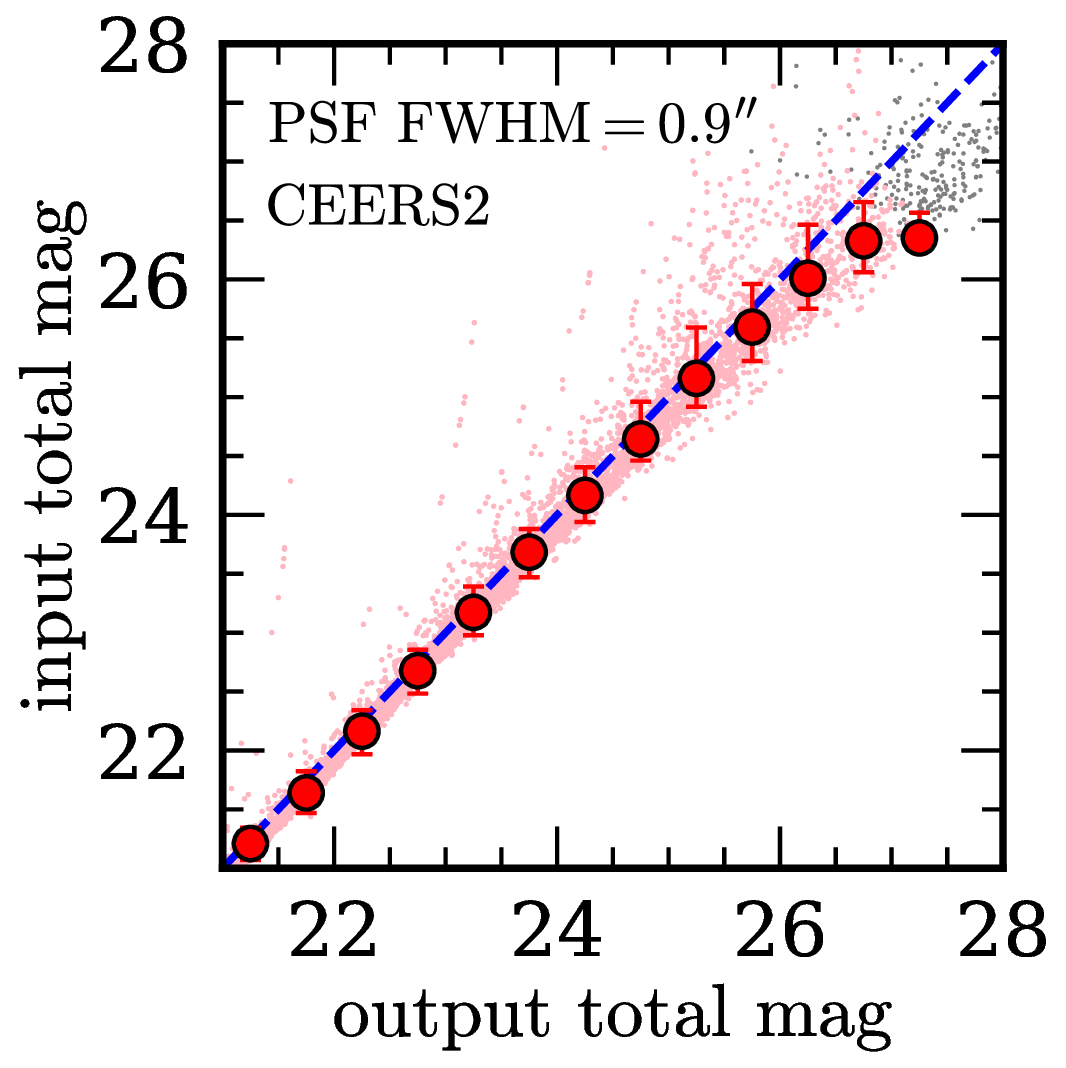}
   \includegraphics[width=0.4\textwidth]{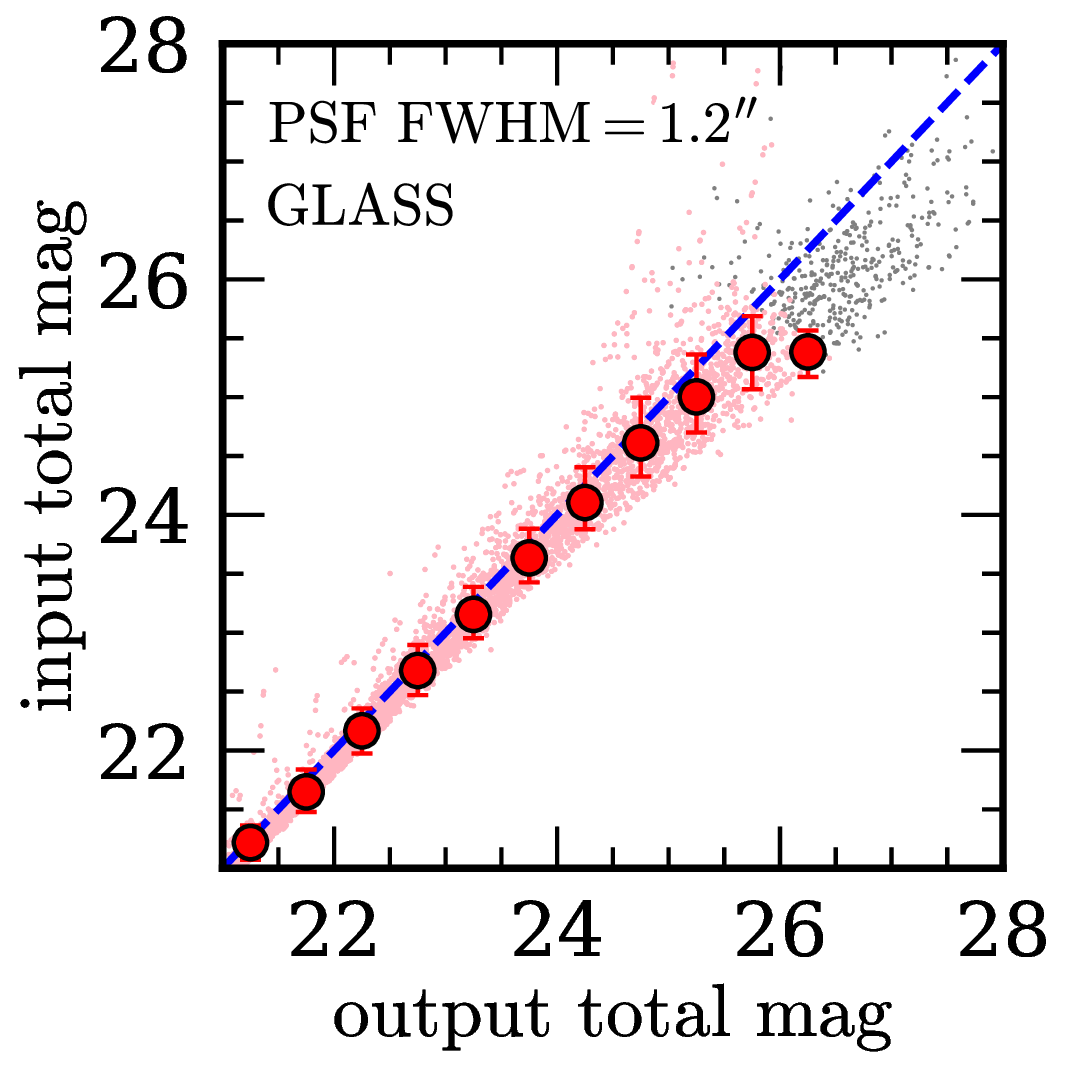}
   \includegraphics[width=0.4\textwidth]{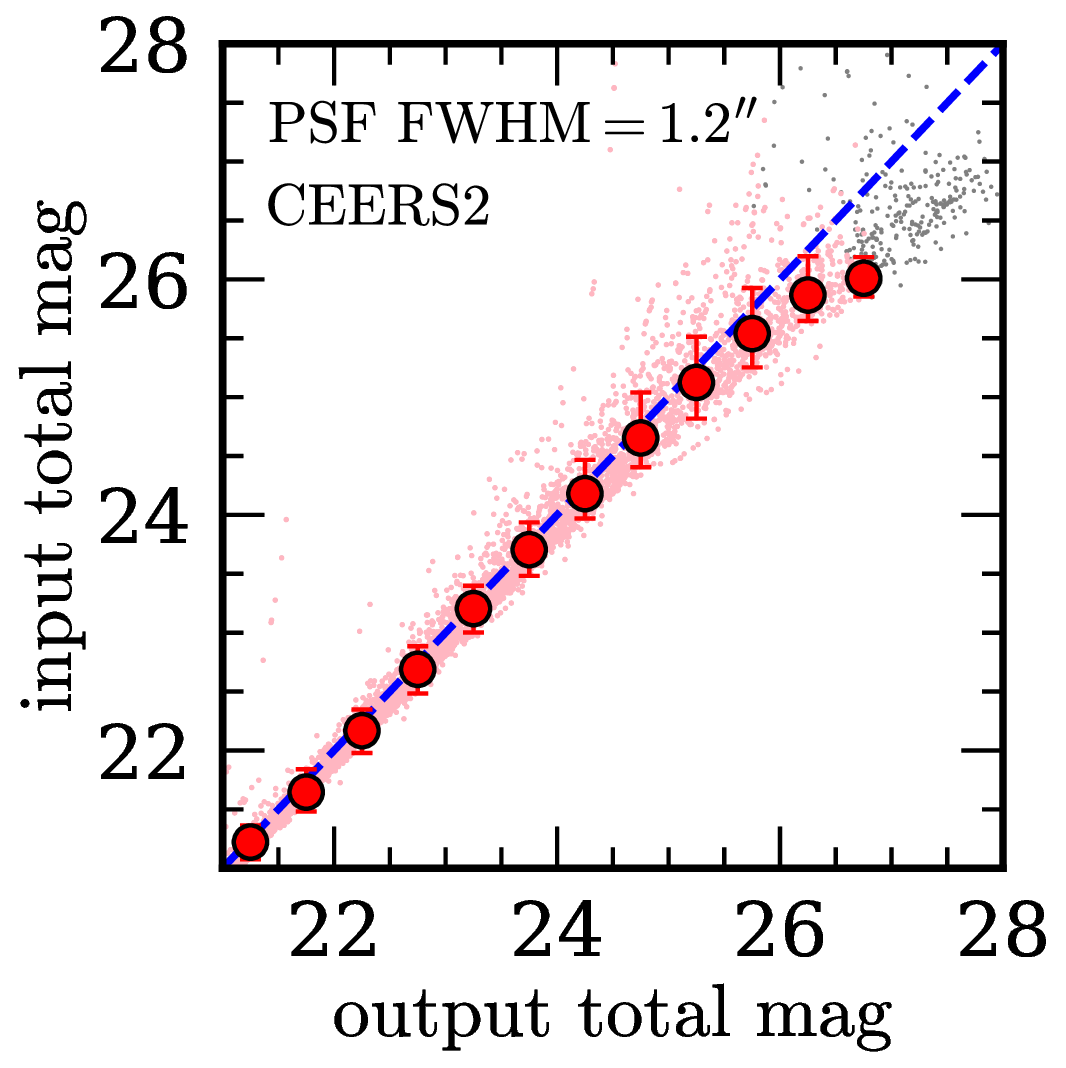}
\caption{
Relation between the input and output total magnitudes for injected PSF sources
in the simulated GREX-PLUS images.
The left panels correspond to the GLASS field and the right panels to the CEERS field.
The top panels show the case with PSF FWHM $=0\farcs9$,
and the bottom panels are PSF FWHM $=1\farcs2$.
Small dots represent individual injected sources:
pink dots indicate sources with aperture magnitudes brighter than the $5\sigma$ limit,
while gray dots indicate sources fainter than the $5\sigma$ limit.
Large red circles indicate, in each magnitude bin, the median of the pink dots,
and the error bars denote the 68th-percentile interval.
The blue dashed line marks equality between the input and output total magnitudes.
}
\label{fig:input_output_mag}
\end{center}
\end{figure*}

As an example, Figure~\ref{fig:simulated_GREX_PLUS_image} presents 
a subregion of the simulated GREX-PLUS images constructed from the JWST NIRCam data in the JADES GOODS-S field.
As expected from the degraded angular resolution, 
bright sources appear more extended 
and show halo-like features due to the PSF wings and the ghost.
In addition, many faint sources are no longer detected 
and are expected to contribute to background fluctuations in apparently blank regions.
As a reference for comparison with a previous infrared space telescope, 
we also show a Spitzer Infrared Array Camera 
(IRAC; \citealt{2004ApJS..154...10F}) $3.6\,\mu$m image 
of the same field (\citealt{2004ApJ...600L..93G}), 
which probes a similar wavelength range 
to those of the GREX-PLUS and JWST images shown here.\footnote{The IRAC image is downloaded 
from \url{https://irsa.ipac.caltech.edu/data/SPITZER/GOODS/index.html}.}
For this IRAC image, we estimate a $5\sigma$ limiting magnitude of 
$24.86$~mag from the standard deviation of fluxes measured in random blank sky apertures
with a diameter of $3\farcs0$, approximately twice the PSF FWHM of the IRAC $3.6\,\mu$m band.
GREX-PLUS is expected to provide images with higher spatial resolution than Spitzer,
mainly owing to its larger primary mirror.

\section{Analysis and Results}
\label{sec:analysis}

\subsection{Source Extraction and Number Counts}
\label{subsec:Source Extraction and Number Counts}

\hspace{0.8em}
We perform source extraction in both the JWST images and the simulated GREX-PLUS images
using SExtractor (\citealt{1996A&AS..117..393B}).
In the following analysis, we focus on the GLASS and CEERS fields,
for which we construct simulated GREX-PLUS images over relatively wide areas
among the deep JWST datasets.
The SExtractor parameter sets adopted for source extraction 
in the JWST and simulated GREX-PLUS images are summarized in Table~\ref{tab:sextractor_params}.
The parameter set for the JWST images is nearly identical to that used in \cite{2023ApJS..265....5H},
except that we adopt slightly lower threshold values to extract sources down to fainter magnitudes.
For the simulated GREX-PLUS images, we use a different set of parameters
to account for the coarser pixel scale and to promote source separation.
Among the extracted sources, we define detected sources as those with aperture magnitudes
brighter than the $5\sigma$ limit.

In this work, we adopt SExtractor as a baseline tool
to quantify how source confusion affects source detection and photometry. 
We note, however, that in strongly confused or blended regimes,
more modern iterative approaches such as PSF-fitting or prior-based deblending
can recover sources more effectively 
(e.g., DAOPHOT for PSF-fitting photometry, \citealt{1987PASP...99..191S}; 
TFIT and T-PHOT for prior-based deblending photometry, 
\citealt{2007PASP..119.1325L}; \citealt{2015A&A...582A..15M}).
Image deconvolution is another potentially useful step to mitigate confusion-induced blending 
when the PSF is stable and accurately characterized
(e.g., demonstrations on JWST MIRI imaging, \citealt{2024AJ....167...96L}).
Prior-based deblending approaches are especially promising in fields with deeper and higher resolution ancillary imaging.
Indeed, such prior-based methods have been successfully demonstrated 
in previous Spitzer deep surveys such as CANDELS and 3D-HST, 
where high resolution HST images are utilized as priors to disentangle severely blended sources 
and extract robust photometry from lower resolution Spitzer IRAC data 
(\citealt{2013ApJS..207...24G}; \citealt{2014ApJS..214...24S}).
For example, the planned Roman High Latitude Wide Area Survey 
with the Roman Space Telescope includes a deep tier 
that will obtain imaging near $2\,\mu$m (\citealt{2025arXiv250510574}), 
which could provide a high resolution prior for deblending and photometry in crowded regions.
Exploring and optimizing these advanced extraction and reconstruction methods is beyond the scope of this paper
and is deferred to future work; therefore, the incompleteness and number counts reported below
should be interpreted as those obtained with our SExtractor-based approach.

Because the extended PSF wings and ghost features around very bright sources 
can hinder the detection of nearby objects and bias local photometry, 
we mask such bright sources and their surrounding regions to mitigate these effects.
The masked areas are determined by visually inspecting each source
with a total magnitude brighter than $21$ mag measured with MAG\_AUTO,
and defining an appropriate mask region for each.
The masked area fraction is $<10$\,\% in the original JWST images,
whereas it increases substantially in the simulated GREX-PLUS images.
For the GLASS field, 
the masked area fraction is $\simeq 36$\,\% for PSF FWHM $=0.9$~arcsec 
and $\simeq 41$\,\% for PSF FWHM $=1.2$~arcsec, 
while for the CEERS field 
it is approximately $31$\,\% and $36$\,\%, respectively.
For detected sources outside the masked regions,
we bin the sources by magnitude and compute the surface number density 
in each magnitude bin to derive the number counts.

\begin{table}
\caption{Median Offsets between Input Total Magnitudes $m_{\rm in}$ 
and SExtractor Output Total Magnitudes $m_{\rm out}$ 
with the $68$ Percentile Ranges based on the Monte Carlo Simulations
with the Simulated GREX-PLUS Images for the GLASS and CEERS Fields}
\begin{center}
{\footnotesize
\begin{tabular}{cccc} \hline
Output Total Mag	& Median $m_{\rm in} - m_{\rm out}$	& $+1\sigma$	& $-1\sigma$ \\ 
(mag)			& (mag)						& (mag) 		& (mag)		\\ \hline
\multicolumn{4}{c}{GLASS with PSF FWHM $=0\farcs9$} \\
$21.32$ & $-0.11$ & $0.14$ & $0.13$ \\
$21.74$ & $-0.11$ & $0.16$ & $0.20$ \\
$22.25$ & $-0.10$ & $0.20$ & $0.17$ \\
$22.75$ & $-0.09$ & $0.20$ & $0.19$ \\
$23.24$ & $-0.10$ & $0.21$ & $0.23$ \\
$23.75$ & $-0.12$ & $0.21$ & $0.24$ \\
$24.24$ & $-0.15$ & $0.21$ & $0.33$ \\
$24.74$ & $-0.16$ & $0.21$ & $0.32$ \\
$25.24$ & $-0.20$ & $0.24$ & $0.43$ \\
$25.74$ & $-0.26$ & $0.35$ & $0.52$ \\
$26.23$ & $-0.44$ & $0.26$ & $0.34$ \\
$26.61$ & $-0.65$ & $0.19$ & $0.33$ \\ \hline
\multicolumn{4}{c}{GLASS with PSF FWHM $=1\farcs2$} \\
$21.30$ & $-0.08$ & $0.15$ & $0.15$ \\
$21.74$ & $-0.09$ & $0.17$ & $0.19$ \\
$22.26$ & $-0.09$ & $0.19$ & $0.19$ \\
$22.76$ & $-0.08$ & $0.21$ & $0.22$ \\
$23.25$ & $-0.10$ & $0.20$ & $0.23$ \\
$23.74$ & $-0.11$ & $0.21$ & $0.25$ \\
$24.24$ & $-0.14$ & $0.23$ & $0.30$ \\
$24.77$ & $-0.16$ & $0.29$ & $0.38$ \\
$25.24$ & $-0.24$ & $0.30$ & $0.36$ \\
$25.68$ & $-0.30$ & $0.31$ & $0.31$ \\
$26.07$ & $-0.69$ & $0.21$ & $0.18$ \\ \hline
\multicolumn{4}{c}{CEERS with PSF FWHM $=0\farcs9$} \\
$21.31$ & $-0.10$ & $0.14$ & $0.14$ \\
$21.74$ & $-0.10$ & $0.17$ & $0.18$ \\
$22.25$ & $-0.09$ & $0.20$ & $0.18$ \\
$22.76$ & $-0.08$ & $0.19$ & $0.18$ \\
$23.25$ & $-0.08$ & $0.19$ & $0.22$ \\
$23.75$ & $-0.07$ & $0.21$ & $0.20$ \\
$24.23$ & $-0.07$ & $0.23$ & $0.24$ \\
$24.74$ & $-0.09$ & $0.18$ & $0.32$ \\
$25.24$ & $-0.08$ & $0.24$ & $0.43$ \\
$25.72$ & $-0.13$ & $0.29$ & $0.37$ \\
$26.23$ & $-0.22$ & $0.26$ & $0.45$ \\
$26.71$ & $-0.38$ & $0.27$ & $0.33$ \\
$27.05$ & $-0.69$ & $0.09$ & $0.21$ \\ \hline
\multicolumn{4}{c}{CEERS with PSF FWHM $=1\farcs2$} \\
$21.30$ & $-0.08$ & $0.15$ & $0.14$ \\
$21.72$ & $-0.07$ & $0.17$ & $0.19$ \\
$22.24$ & $-0.07$ & $0.19$ & $0.18$ \\
$22.75$ & $-0.06$ & $0.20$ & $0.20$ \\
$23.25$ & $-0.04$ & $0.20$ & $0.19$ \\
$23.73$ & $-0.03$ & $0.22$ & $0.23$ \\
$24.24$ & $-0.06$ & $0.21$ & $0.29$ \\
$24.73$ & $-0.08$ & $0.25$ & $0.39$ \\
$25.23$ & $-0.11$ & $0.31$ & $0.39$ \\
$25.73$ & $-0.19$ & $0.29$ & $0.39$ \\
$26.23$ & $-0.36$ & $0.22$ & $0.33$ \\
$26.62$ & $-0.61$ & $0.16$ & $0.18$ \\ \hline
\end{tabular}
}
\end{center}
\label{tab:output_input_total_mag_MC}
\end{table}

\begin{table}
\caption{Detection Completeness as a Function of Input Total Magnitude
for the Two Cases with PSF FWHM $=0\farcs9$ and $1\farcs2$ 
based on the Monte Carlo Simulations with the Simulated GREX-PLUS Images 
for the GLASS and CEERS Fields}
\begin{center}
\begin{tabular}{ccc} \hline
Input Total Mag		&  \multicolumn{2}{c}{Detection Completeness} \\ 
(mag)			&  (PSF FWHM = $0\farcs9$)	& (PSF FWHM = $1\farcs2$) \\\hline
\multicolumn{3}{c}{GLASS}  \\
$21.25$ & $0.92$ & $0.88$ \\
$21.75$ & $0.90$ & $0.87$ \\
$22.25$ & $0.88$ & $0.84$ \\
$22.75$ & $0.87$ & $0.81$ \\
$23.25$ & $0.82$ & $0.77$ \\
$23.75$ & $0.78$ & $0.73$ \\
$24.25$ & $0.69$ & $0.64$ \\
$24.75$ & $0.64$ & $0.63$ \\
$25.25$ & $0.56$ & $0.50$ \\
$25.75$ & $0.50$ & $0.16$ \\
$26.25$ & $0.22$ & $0.02$ \\
$26.75$ & $0.04$ & $0.01$ \\
$27.25$ & $0.02$ & $0.01$ \\
\hline
\multicolumn{3}{c}{CEERS} \\
$21.25$ & $0.92$ & $0.93$ \\
$21.75$ & $0.93$ & $0.89$ \\
$22.25$ & $0.90$ & $0.87$ \\
$22.75$ & $0.86$ & $0.85$ \\
$23.25$ & $0.82$ & $0.79$ \\
$23.75$ & $0.77$ & $0.67$ \\
$24.25$ & $0.71$ & $0.67$ \\
$24.75$ & $0.66$ & $0.54$ \\
$25.25$ & $0.61$ & $0.52$ \\
$25.75$ & $0.52$ & $0.47$ \\
$26.25$ & $0.41$ & $0.18$ \\
$26.75$ & $0.14$ & $0.04$ \\
$27.25$ & $0.02$ & $0.01$ \\
\hline
\end{tabular}
\end{center}
\label{tab:detection_completeness}
\end{table}

The obtained number counts are shown in Figure~\ref{fig:number_counts_GLASS}.
Because sources brighter than $21$~mag in total magnitude are masked,
we do not show number counts in that bright regime.
The left panels present the results for the GLASS field. 
The top left panel corresponds to PSF FWHM $=0.9$~arcsec and the bottom left panel to PSF FWHM $=1.2$~arcsec.
In the bright magnitude range of $\simeq21$--$24$~mag,
the number counts derived from the JWST images 
and those from the simulated GREX-PLUS images agree well.
However, toward fainter magnitudes of $\gtrsim24$~mag,
the number counts measured from the simulated GREX-PLUS images decrease,
and the discrepancy relative to the JWST-based number counts becomes larger.
This behavior marks the onset of confusion-induced blending, 
which reduces the source detection completeness,
as quantified by the Monte Carlo completeness analysis in Section \ref{subsec:Detection Completeness}.
We note that the masking of bright sources mitigates additional blending 
and structured background in their vicinity; 
without such masking, the effective incompleteness would set in at brighter magnitudes.
The PSF FWHM $=1.2$~arcsec case shows a similar trend, 
but the onset of incompleteness occurs at slightly brighter magnitudes.
The results for the CEERS field are presented in the right panels 
and show the same overall behavior.
Because the simulated GREX-PLUS images for the CEERS field are slightly deeper, 
the drop in the number counts relative to the JWST-based ones at the faint end 
is more gradual than in the GLASS field.

In the next subsection, 
we quantify the detection completeness based on Monte Carlo simulations.
We then present completeness-corrected number counts in 
Section~\ref{subsec:Completeness-corrected Number Counts}.

\begin{figure*}
\begin{center}
   \includegraphics[width=0.4\textwidth]{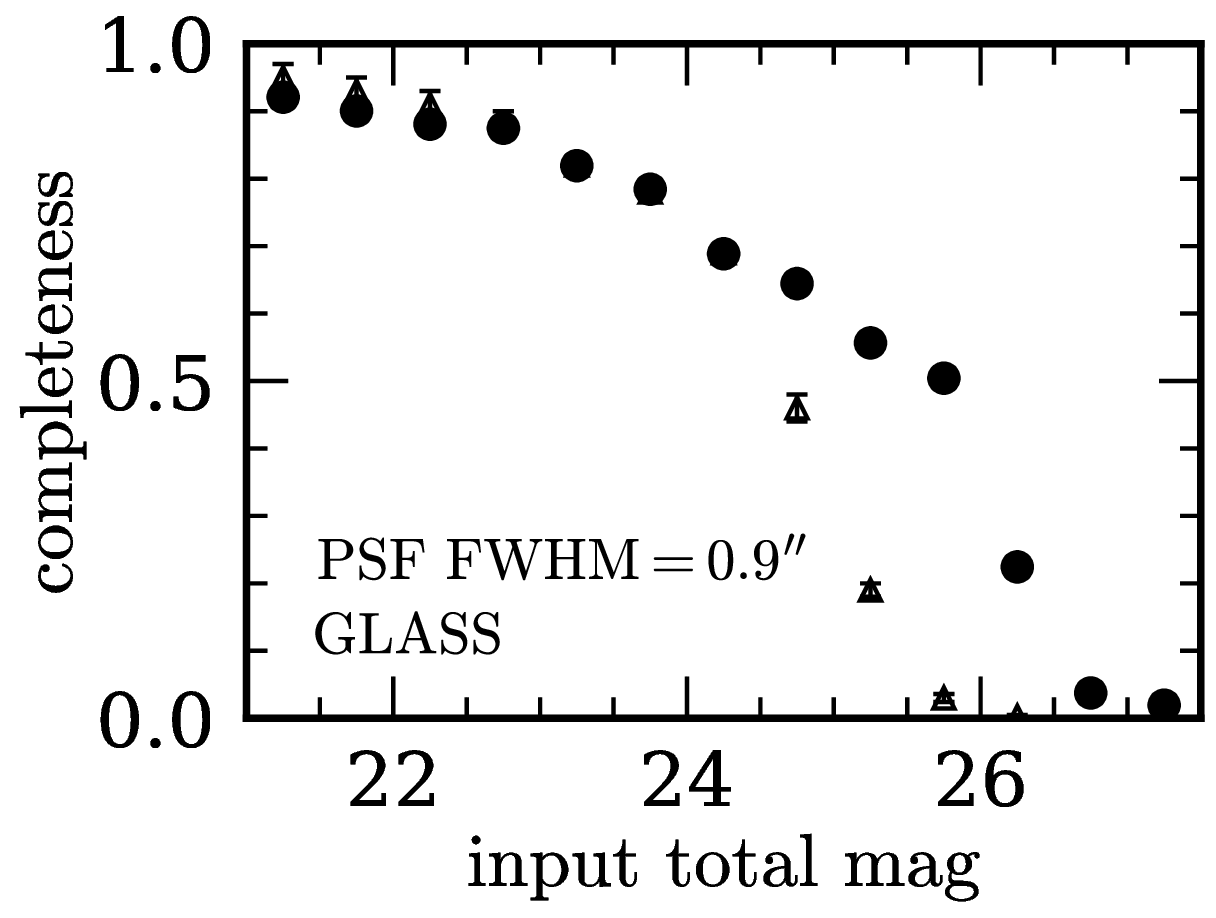}
   \includegraphics[width=0.4\textwidth]{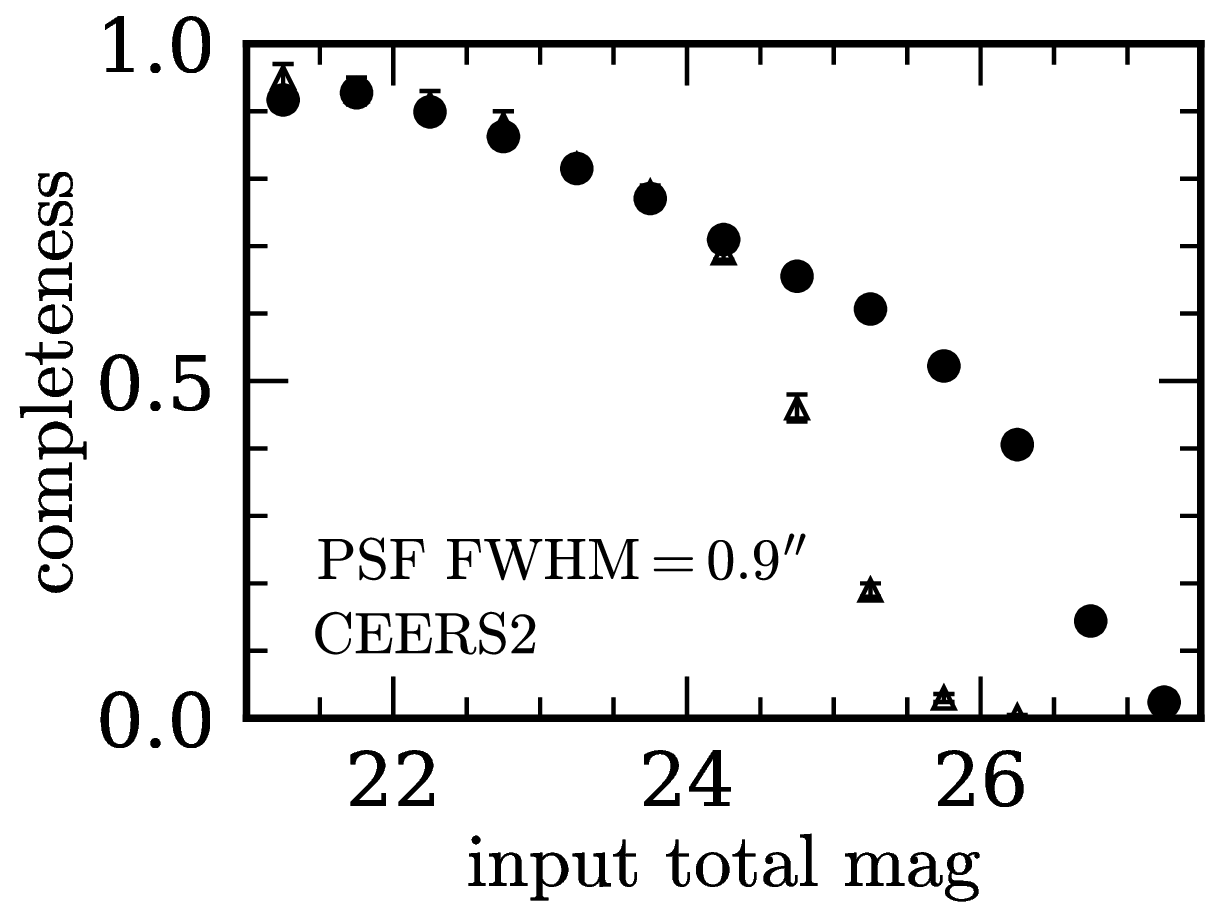}
   \includegraphics[width=0.4\textwidth]{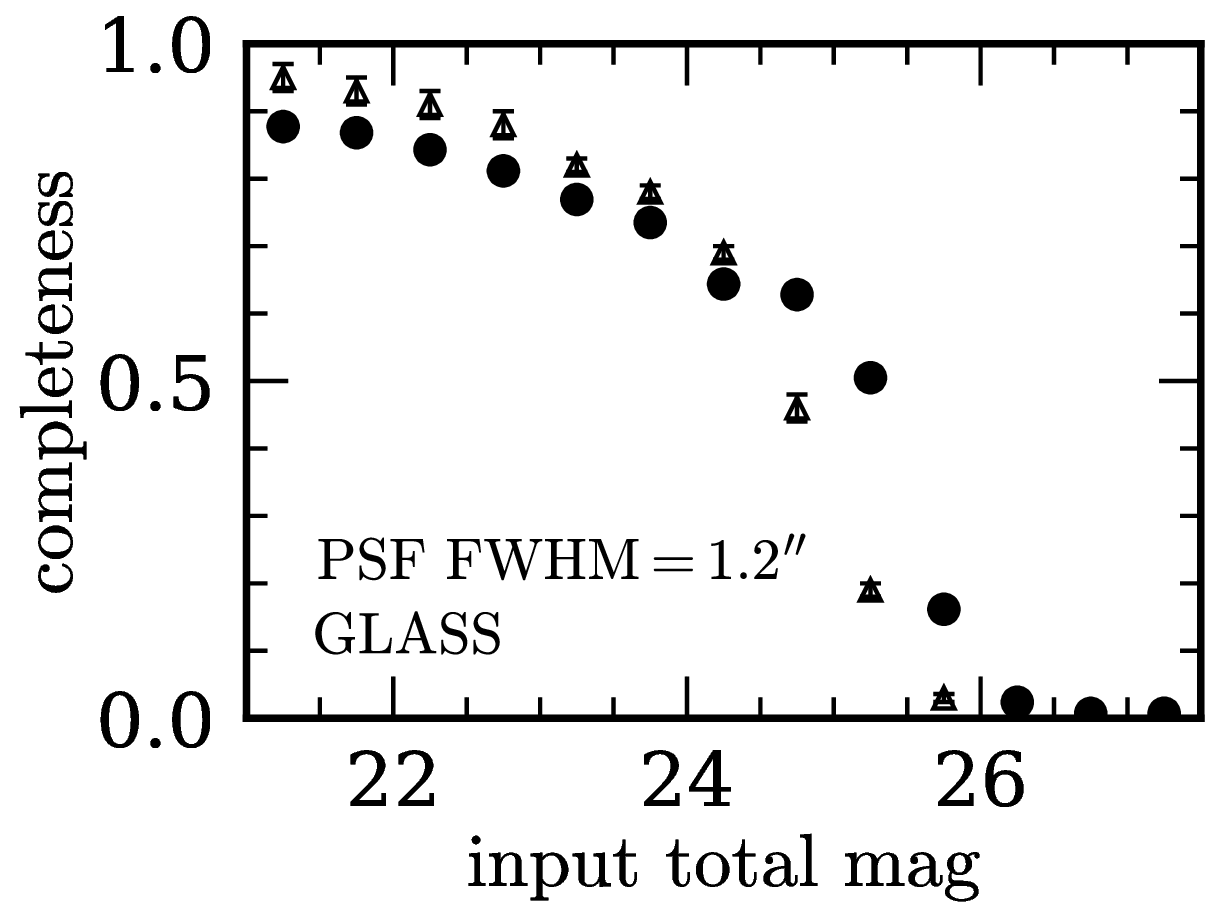}
   \includegraphics[width=0.4\textwidth]{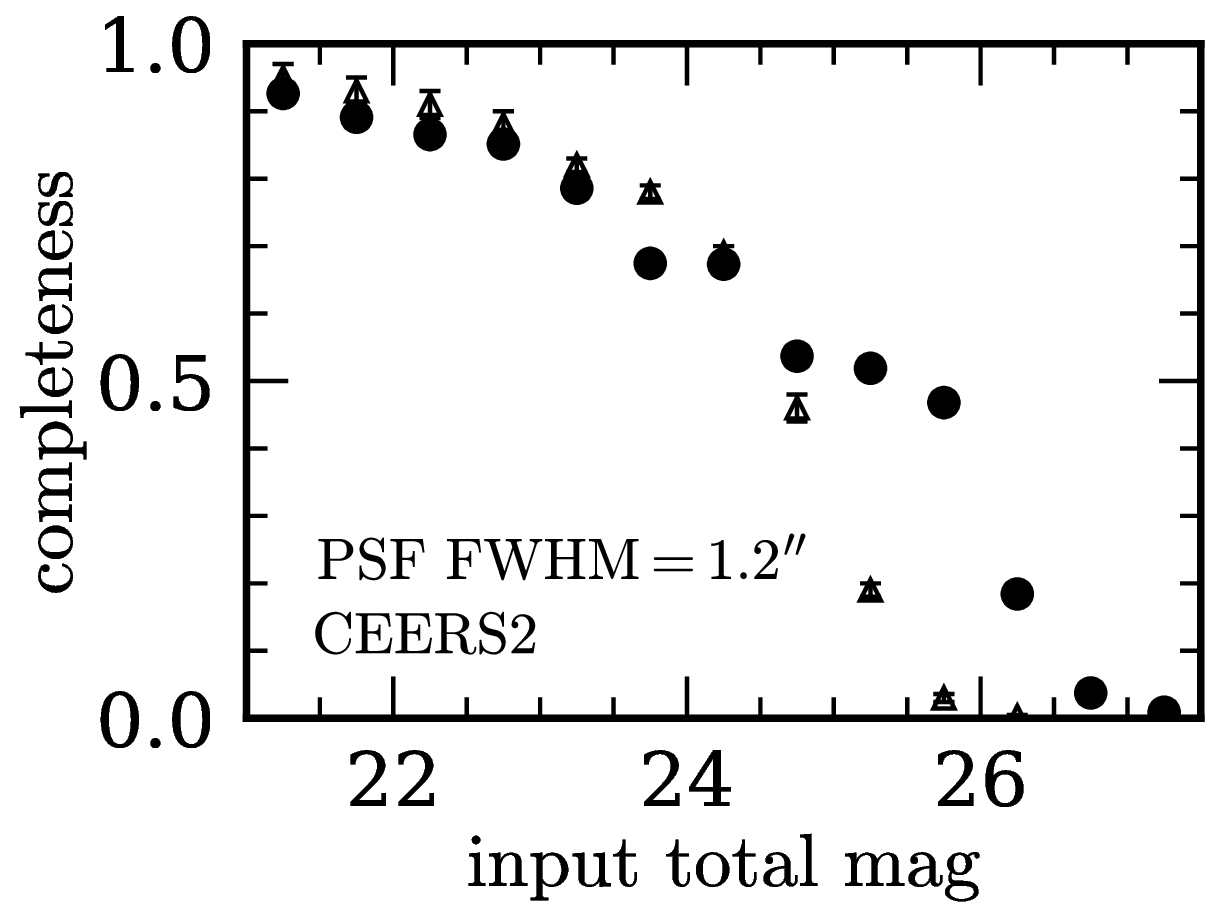}
\caption{
Detection completeness as a function of input total magnitude.
Black circles indicate the completeness estimated from our Monte Carlo simulations
based on the simulated GREX-PLUS images.
Black triangles indicate the completeness derived from the Spitzer SEDS images
(\citealt{2013ApJ...769...80A}).
The left panels denote the GLASS field and the right panels are the CEERS field.
The top panels correspond to the case with PSF FWHM $=0\farcs9$,
and the bottom panels to PSF FWHM $=1\farcs2$.
}
\label{fig:completeness}
\end{center}
\end{figure*}

\subsection{Detection Completeness}
\label{subsec:Detection Completeness}

\hspace{0.8em}
To quantitatively estimate the detection completeness, we perform Monte Carlo simulations.
We inject 10,000 artificial sources with input total magnitudes spanning $21$--$28$~mag
into the simulated GREX-PLUS images and attempt to recover them with SExtractor 
adopting the same source extraction configuration as used in Section \ref{subsec:Source Extraction and Number Counts}.
We define the detection completeness as the fraction of injected sources 
that are extracted by SExtractor 
and have aperture magnitudes brighter than the $5\sigma$ limit.
Note that the detection completeness estimated here 
is specific to our SExtractor-based approach,
and it may improve with iterative PSF-fitting or deblending methods in confused regions. 
Because most sources are unresolved and can be treated as point sources,
particularly the relatively faint ones near the detection limit where incompleteness becomes significant,
we assume the PSF$+$ghost profile for the injected point sources. 
When injecting sources, we normalize the PSF$+$ghost image 
so that its total flux matches the input total magnitude.
Injecting many sources into a single image would artificially enhance source confusion 
and lead to an underestimation of the completeness.
We therefore select random positions from regions that are not masked in the number count analysis, 
extract $20'' \times 20''$ cutouts, 
and inject a PSF$+$ghost source at the center of each cutout.
We then run SExtractor on these images and compile a catalog of recovered sources
that are detected within $\pm0.5$~pixel of the input position
(i.e., within $0.25$~arcsec).
For these recovered sources, we examine the output total magnitudes 
measured as MAG\_AUTO with SExtractor,
and compare them with the input total magnitudes.
Finally, we bin the injected sources by their input total magnitude 
and compute, in each bin, the fraction of sources that are recovered 
and have aperture magnitudes brighter than the $5\sigma$ limit.  
We adopt this fraction as the detection completeness.

Figure~\ref{fig:input_output_mag} compares the input total magnitudes with the output total magnitudes
measured as MAG\_AUTO values using SExtractor.
The results for the GLASS field are presented in the left panels.
In the top left panel with PSF FWHM $=0.9$~arcsec,
the output magnitudes are consistent with the input magnitudes within the $1\sigma$ scatter from $\simeq21$ to $26$~mag, 
although the median output magnitude is systematically fainter by about $0.1$~mag.
Toward fainter magnitudes of $\gtrsim26$~mag,
the output total magnitudes become systematically fainter than the input values.
This is probably because, for fainter sources, 
the low surface brightness outskirts are more easily buried in the background, 
causing MAG\_AUTO to miss part of the extended light and underestimate the total flux.
The bottom left panel for PSF FWHM $=1.2$~arcsec shows a similar trend,
but the deviation from the one-to-one relation of input $=$ output
begins at magnitudes that are about $0.5$~mag brighter.
The results for the CEERS field are presented in the right panels and show basically the same general trend.
Because the CEERS images are slightly deeper, the input and output magnitudes remain in good agreement
to magnitudes that are approximately $0.5$~mag fainter than in the GLASS field.
We summarize, for each magnitude bin, the median offset between the input and output total magnitudes
and its 68th-percentile range in Table~\ref{tab:output_input_total_mag_MC}.
Because the output total magnitudes of detected sources are on average biased faint,
we apply a correction by shifting the measured magnitudes brighter by the median input$-$output offset.

Figure~\ref{fig:completeness} shows the detection completeness as a function of input total magnitude,
and the corresponding values are listed in Table~\ref{tab:detection_completeness}.
The results for the GLASS field are presented in the left panels.
In the top left panel with PSF FWHM $=0.9$~arcsec,
the completeness exceeds $90$\,\% for bright sources around $21$~mag,
but gradually decreases toward fainter magnitudes,
reaching about $50$\,\% at $\simeq25.5$~mag and dropping to nearly zero by around $26.5$~mag.
The completeness does not reach unity 
even in bright bins that are far above the $5\sigma$ depth.
This is unlikely to be driven by sensitivity, 
and instead suggests that confusion-induced blending already affects source recovery in the simulated images.

For comparison, we also plot the completeness measured for the
Spitzer Extended Deep Survey (SEDS) images from \cite{2013ApJ...769...80A}, 
which provide deep imaging at a similar wavelength. 
Overall, the completeness trends as a function of input magnitude are broadly similar
between our results and those of \cite{2013ApJ...769...80A}.
In detail, because the simulated GREX-PLUS image is deeper,
its completeness remains higher down to fainter magnitudes.
It is also worth noting that 
the SEDS completeness rises somewhat more rapidly toward brighter magnitudes. 
In practice, it begins to increase at input total magnitudes roughly $1$~mag brighter,
yet it catches up to the simulated GREX-PLUS completeness curve.
This difference is plausibly attributable to the source detection methodology.
Spitzer images can be somewhat more affected by source confusion than GREX-PLUS,
because the Spitzer primary mirror diameter is $0.85$~m, smaller than that of GREX-PLUS,
leading to a modestly lower angular resolution and hence more frequent blending between neighboring sources.
Accordingly, \cite{2013ApJ...769...80A} have adopted PSF-fitting photometry using
StarFinder (\citealt{2000SPIE.4007..879D}),
iteratively fitting the PSF to detect sources even in crowded regions 
(see also, \citealt{2018ApJS..237...39A}).
This approach can achieve higher completeness than our simple SExtractor-based detection.

The results for the CEERS field are shown in the right panels and exhibit the same overall behavior.
Because the CEERS images are slightly deeper, the completeness curves are shifted 
toward fainter magnitudes by about $0.5$~mag relative to the GLASS field. 
For the case with PSF FWHM $=1.2$~arcsec, the completeness is shifted to brighter magnitudes
relative to the $0.9$~arcsec case in both fields, reflecting the shallower effective depth,
while the overall trend remains similar.

\begin{figure*}
\begin{center}
   \includegraphics[width=0.4\textwidth]{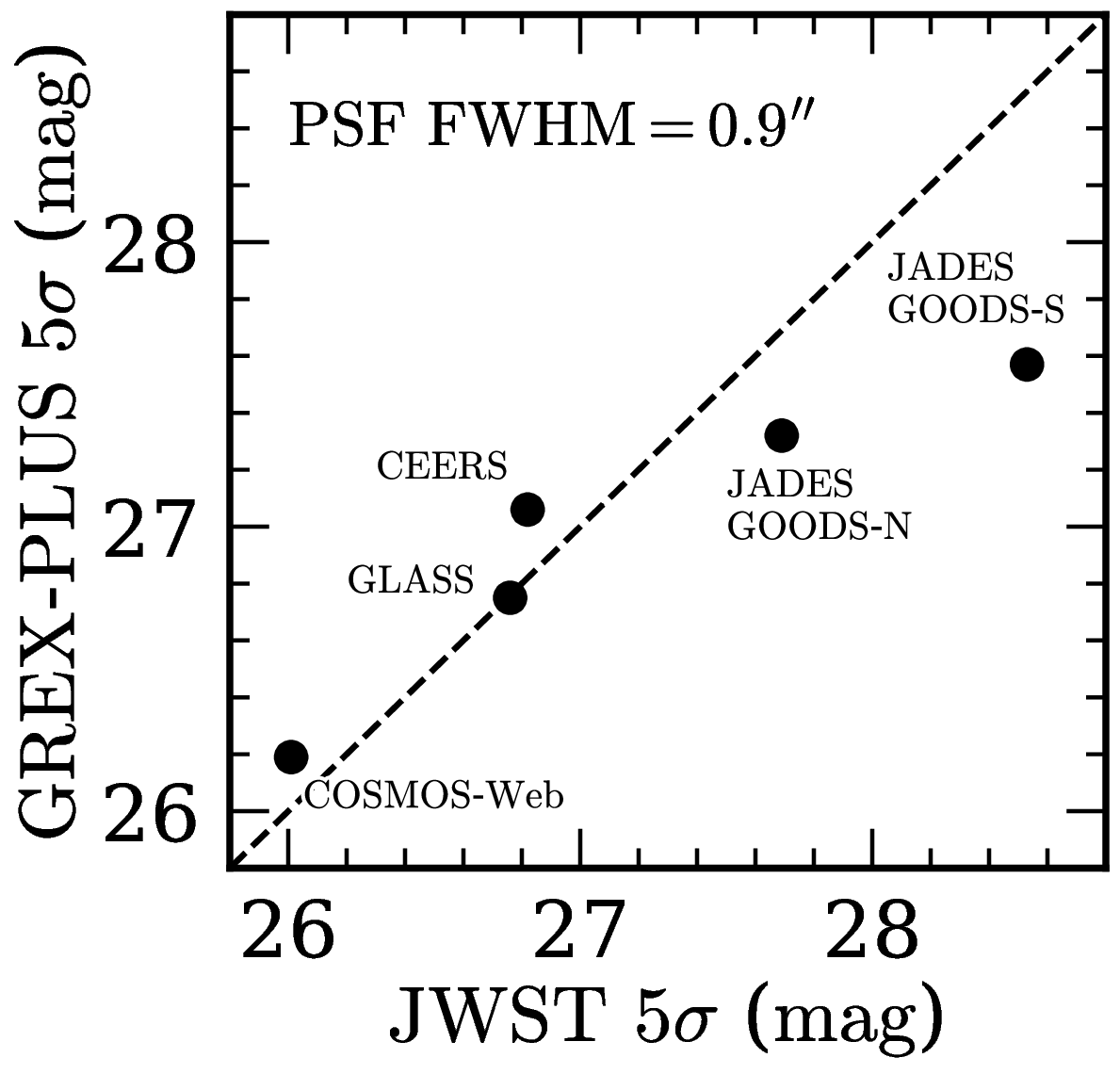}
\hspace{1em}
   \includegraphics[width=0.4\textwidth]{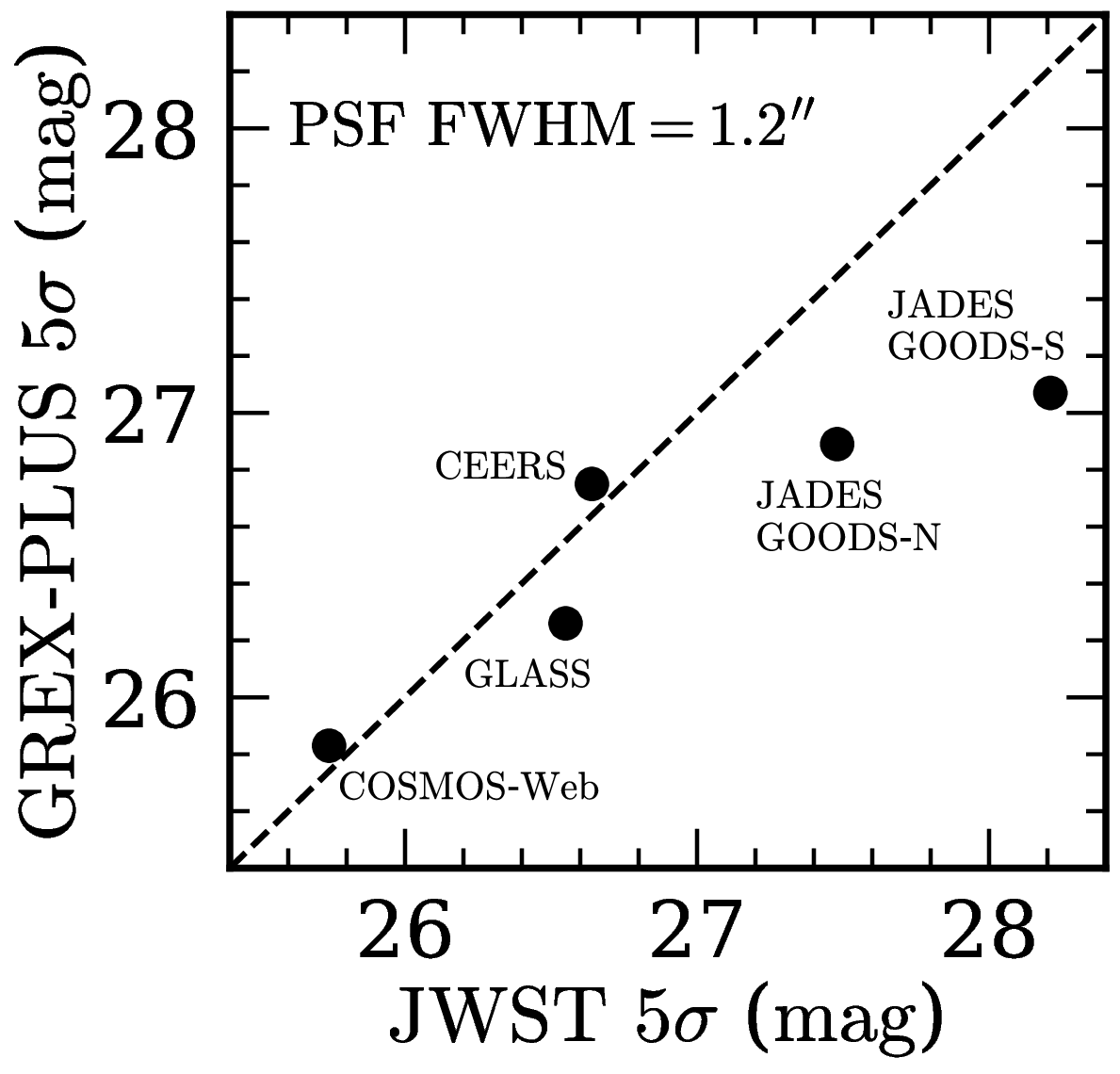}
\caption{Comparison of the $5\sigma$ limiting magnitudes 
measured from the simulated GREX-PLUS images and the original JWST images. 
The left panel corresponds to the case in which the convolution kernel 
used to generate the simulated GREX-PLUS images has PSF FWHM $=0\farcs9$,  
while the right panel shows the case with PSF FWHM $=1\farcs2$. 
To ensure a direct comparison, the JWST limiting magnitudes on the horizontal axes 
are measured using the same aperture sizes as the GREX-PLUS images 
($2\farcs0$ for the left panel and $2\farcs4$ for the right panel). 
The dashed line indicates the one-to-one relation representing the expected depth 
if both datasets have identical background noise levels within the matched apertures.
The deeper data points tend to lie below the dashed line, 
suggesting that source confusion effects introduced by the PSF$+$ghost convolution 
become more significant for deeper data.
}
\label{fig:limitmag_rev}
\end{center}
\end{figure*}

\section{Discussion}
\label{sec:results_and_discussion}

\subsection{Limiting Magnitude Comparison}
\label{subsec:Limiting Magnitude Comparison}

\hspace{0.8em}
In Section~\ref{sec:Simulated Imaging Data for GREX-PLUS}, 
we have estimated the limiting magnitudes 
for both the simulated GREX-PLUS images and the original JWST images
by performing random aperture photometry in regions with no detected sources
and deriving the standard deviations from the distribution of the measured aperture fluxes.
Figure~\ref{fig:limitmag_rev} compares the obtained limiting magnitudes.
To ensure a direct and consistent comparison, 
we match the aperture sizes used for both datasets:
the $2\farcs0$ aperture limiting magnitudes for the JWST images are compared 
with the PSF FWHM $=0.9$~arcsec cases,
and the $2\farcs4$ limiting magnitudes are compared 
with the PSF FWHM $=1.2$~arcsec cases.
The JWST limiting magnitudes span a wide range from COSMOS-Web to JADES,
which effectively corresponds to probing simulated GREX-PLUS images 
over a range of integration times.

In Figure~\ref{fig:limitmag_rev}, 
the dashed line represents a one-to-one relation, 
which is expected if the two datasets have identical background noise levels 
within the matched apertures. 
As can be seen, for COSMOS-Web, GLASS, and CEERS, 
the simulated GREX-PLUS depths are broadly consistent with the JWST depths, 
whereas for the deeper JADES images 
the simulated GREX-PLUS images become noticeably shallower.
Since the aperture sizes are identical,
this is likely driven by the convolution with the GREX-PLUS PSF$+$ghost kernel.
This process redistributes flux from unresolved faint sources and bright objects across the image,
thereby increasing the apparent background fluctuations measured in nominally blank regions.
In particular, this excess noise includes confusion noise from the superposition of unresolved sources, 
as well as additional contributions from the extended PSF wings and ghost halos of bright sources.
The larger offsets from the dashed line for PSF FWHM $=1.2$~arcsec than for $0.9$~arcsec
suggest that this confusion-induced noise increases systematically as the effective PSF broadens.

\begin{figure*}
\begin{center}
   \includegraphics[width=0.505\textwidth]{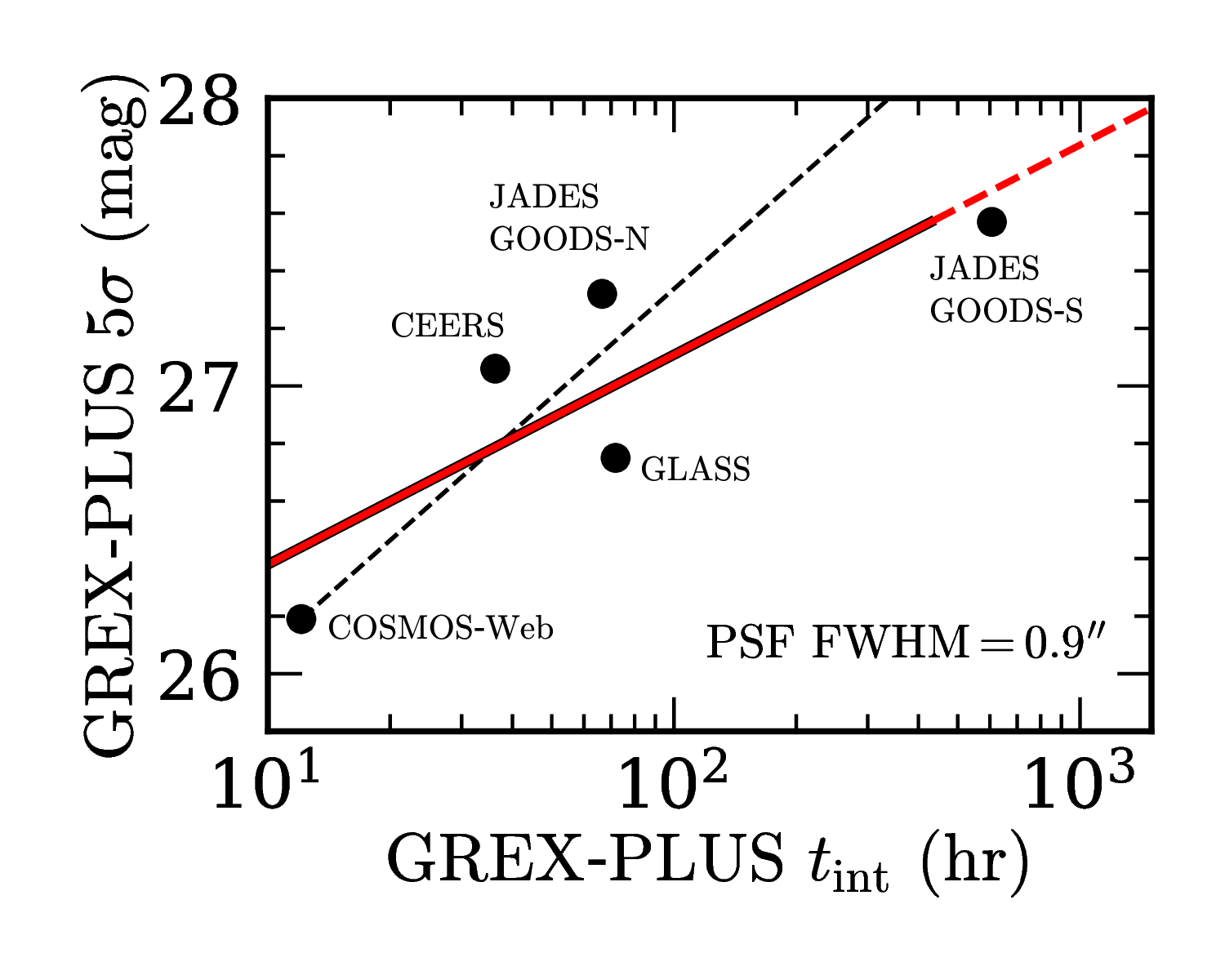}
   \includegraphics[width=0.485\textwidth]{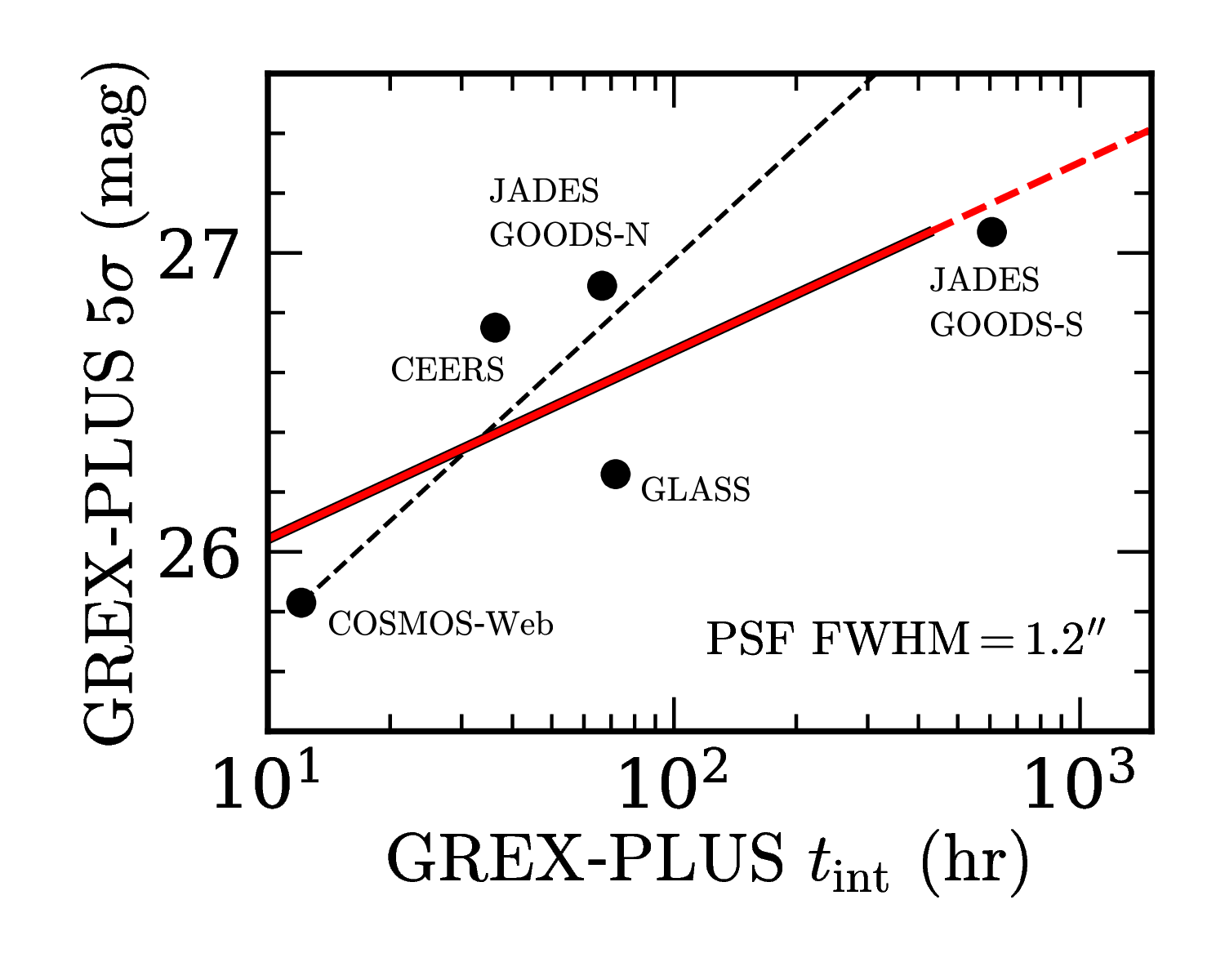}
\caption{
The $5 \sigma$ limiting magnitudes measured in simulated GREX-PLUS images 
as a function of the GREX-PLUS integration time required to reach the corresponding depth, 
estimated by scaling the integration times of the underlying JWST images. 
The left panel shows the case in which the images are convolved with 
a PSF$+$ghost kernel with PSF FWHM $=0\farcs9$, 
while the right panel shows the case with PSF FWHM $=1\farcs2$. 
The red solid line indicates the relation obtained from linear regression fits to the data points.
The red dashed line shows the extrapolation beyond the deepest data point,
where the actual slope is expected to become shallower.
The black dashed line corresponds to the relation expected for square-root improvement with integration time, 
normalized to the COSMOS-Web data point.
}
\label{fig:limitmag_texp}
\end{center}
\end{figure*}

An important point is that the limiting magnitudes of the simulated GREX-PLUS images  
estimated from blank sky aperture fluctuations continue to improve 
as the limiting magnitudes of the JWST images become deeper, 
with no clear evidence for a sharp plateau down to about $27$~mag.
This implies that, over the depth range probed here, 
the blank sky aperture noise metric does not yet appear to be limited by a confusion floor,
i.e., it shows no clear sign of having reached the confusion limit.
At the same time, our source extraction results indicate that
confusion-induced blending reduces detection completeness at significantly brighter magnitudes. 
This incompleteness becomes apparent at around $24$~mag and increases toward fainter levels.
It should also be noted that the offset from the dashed line increases toward deeper data. 
This implies that the depth improvement with integration time becomes progressively less efficient.
When planning the GREX-PLUS survey, it is therefore necessary to account for this effect
when estimating the required integration times.

\begin{figure*}
\begin{center}
   \includegraphics[height=0.4\textwidth]{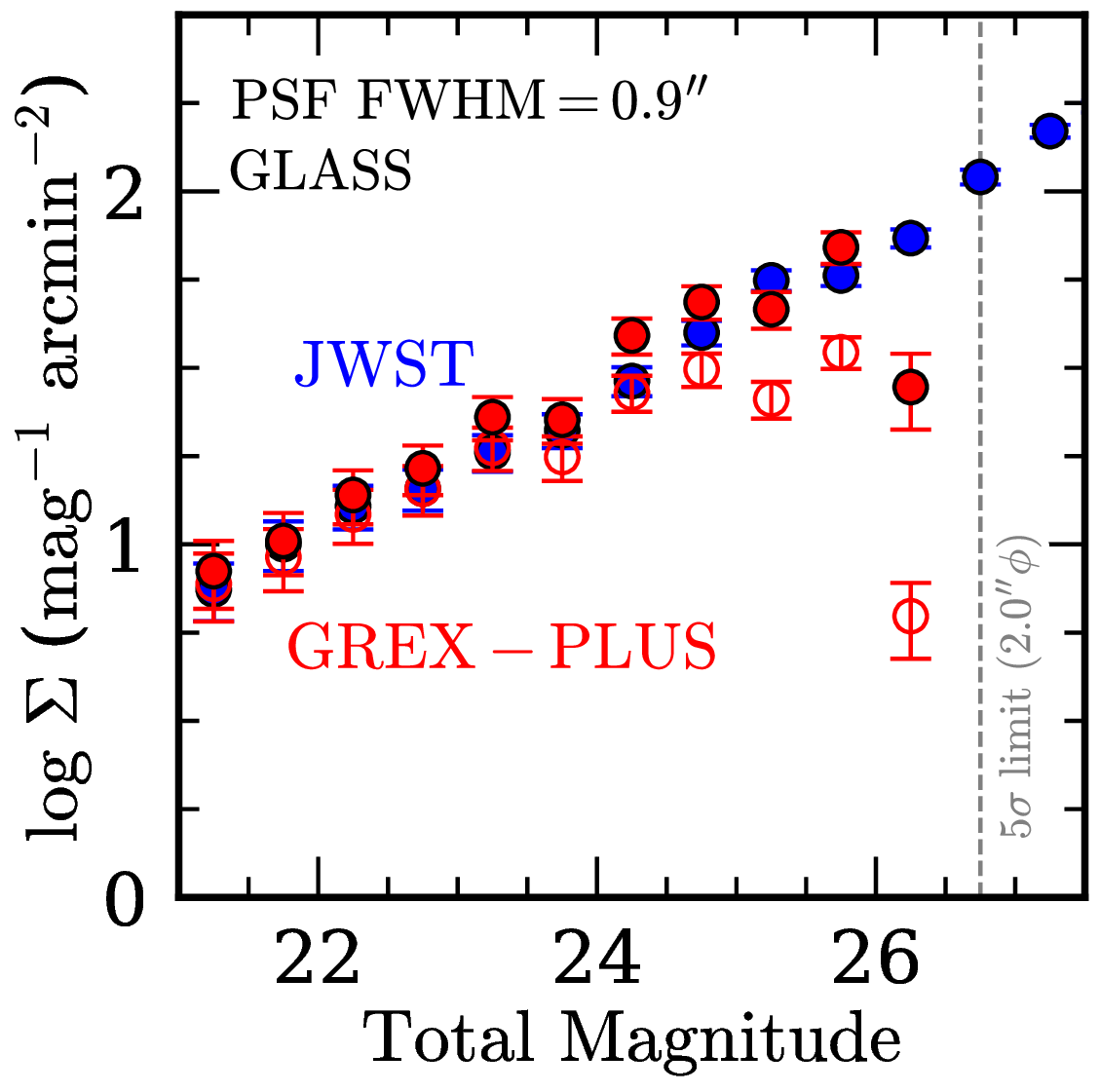}
   \includegraphics[height=0.4\textwidth]{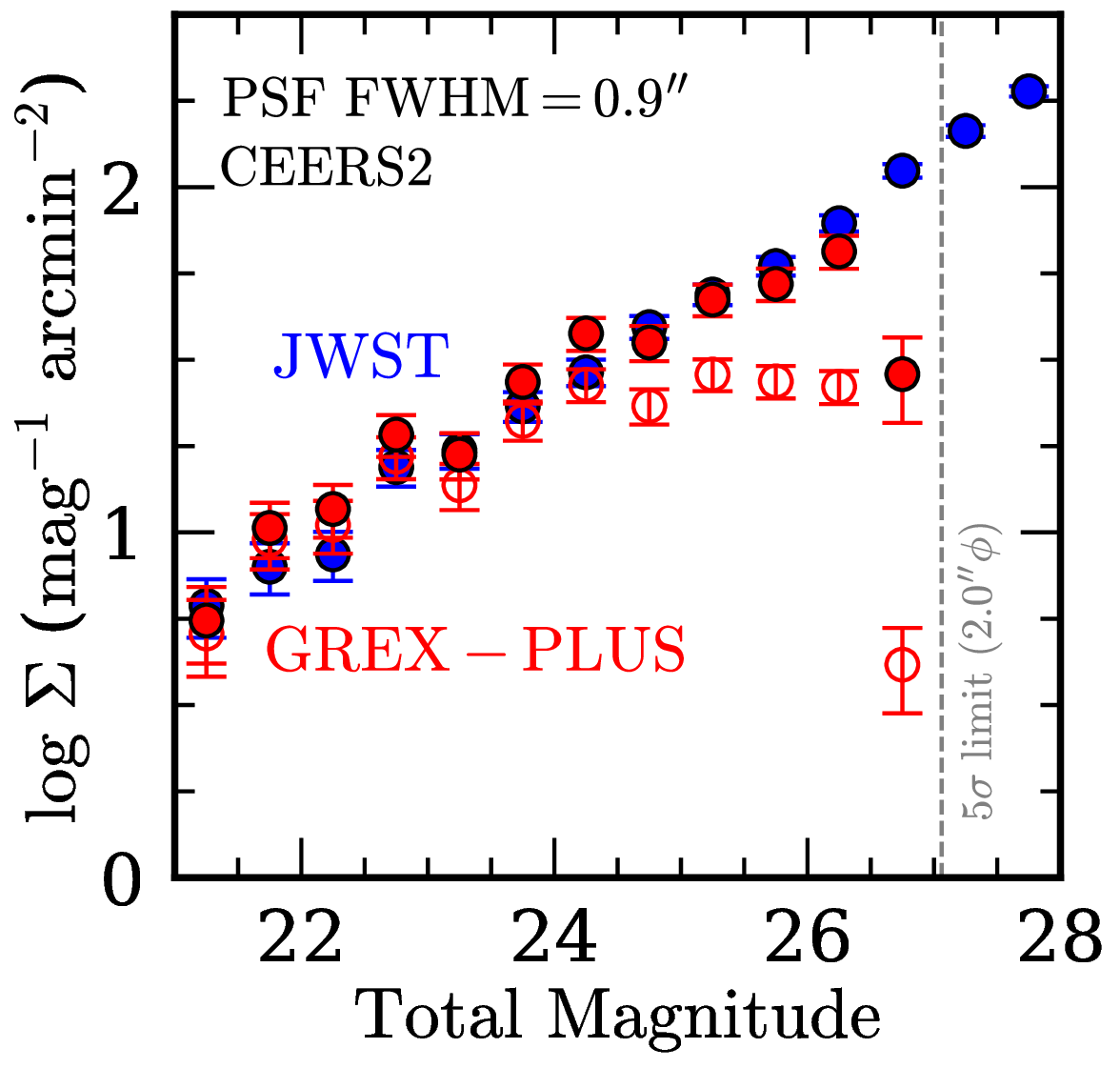}
   \includegraphics[height=0.4\textwidth]{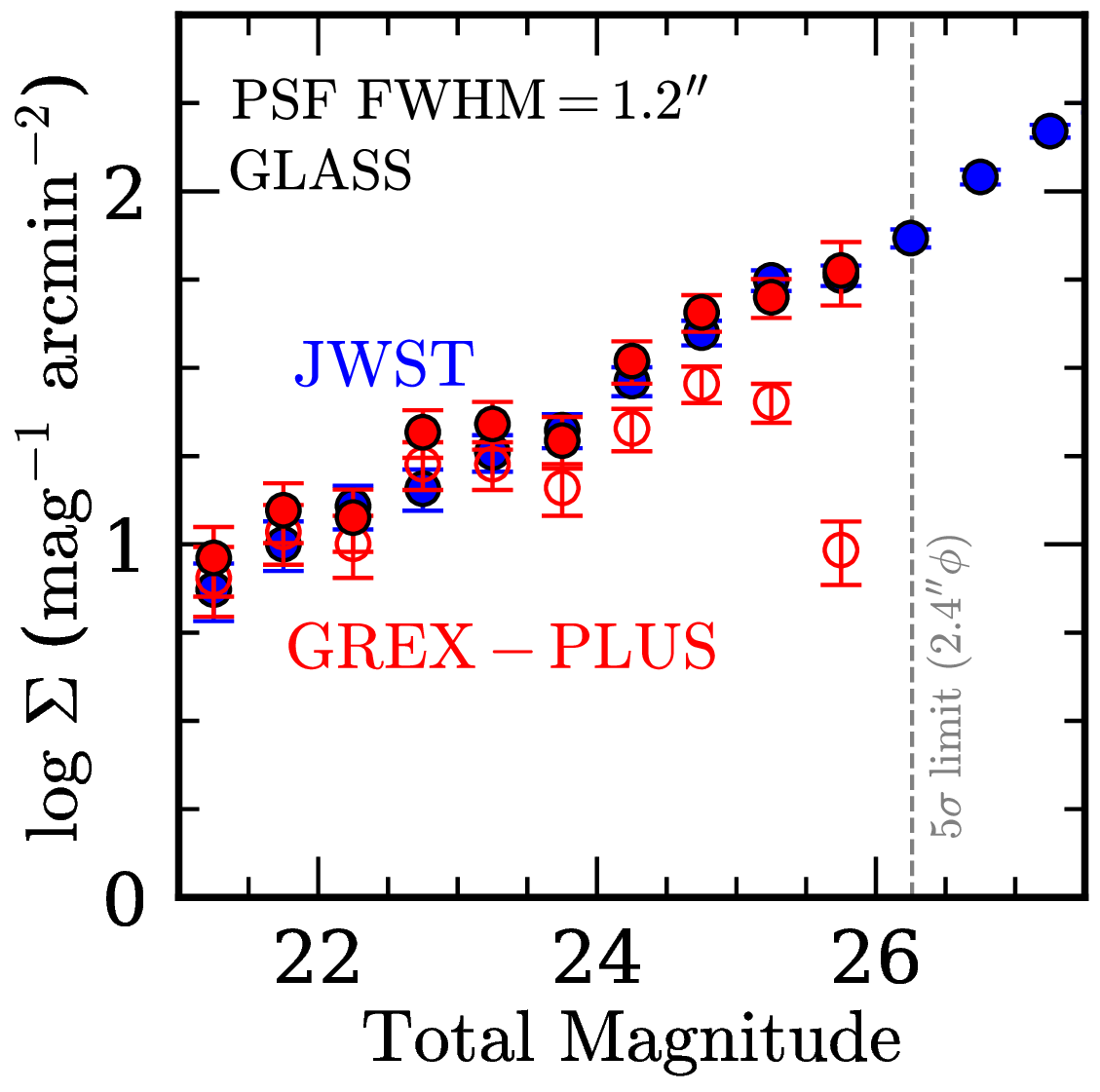}
   \includegraphics[height=0.4\textwidth]{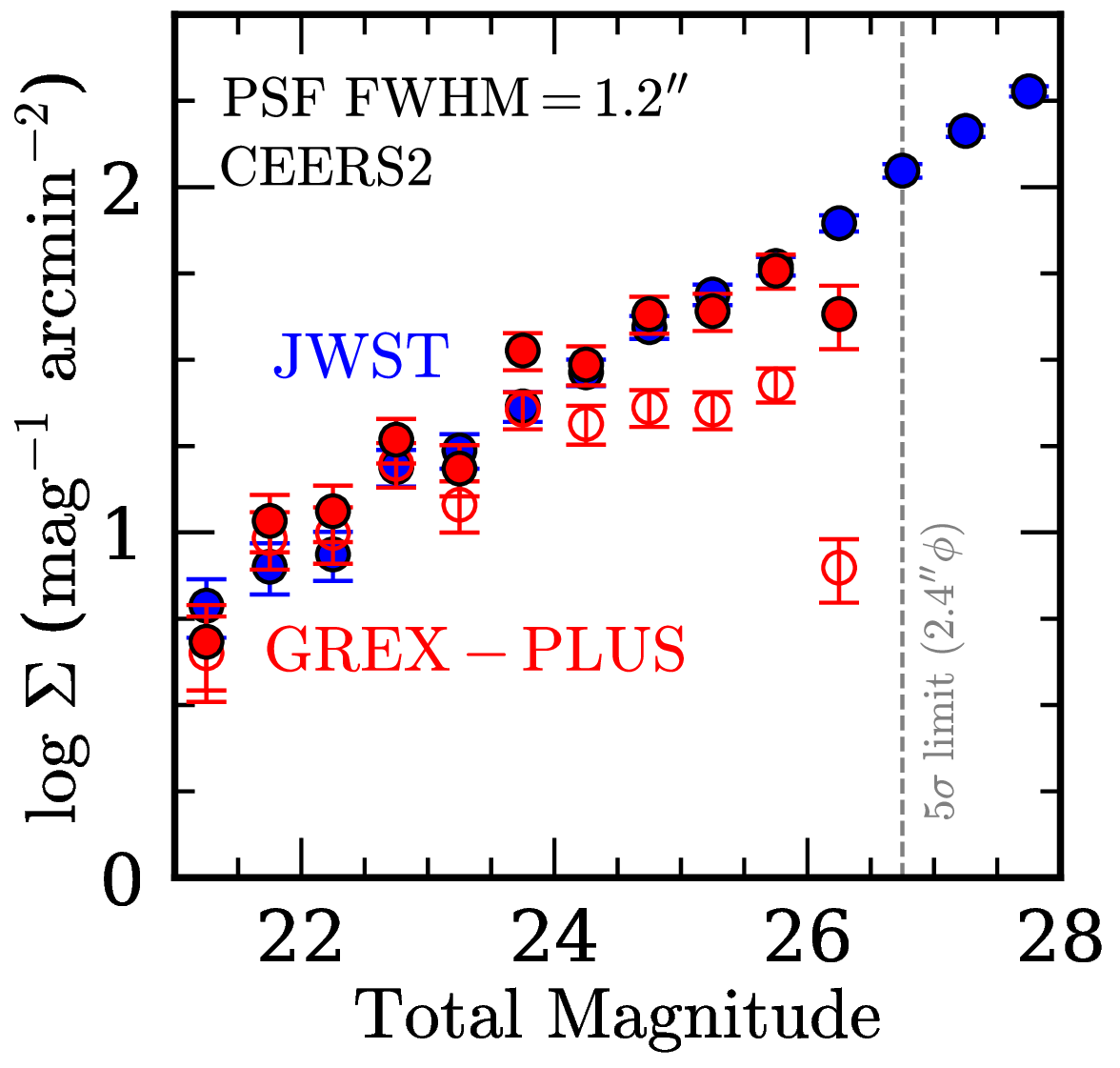}
\caption{
Number counts corrected for the output total magnitude offset and detection completeness
based on the Monte Carlo simulations.
The left panels correspond to the GLASS field and the right panels to the CEERS field.
The top panels denote the case with PSF FWHM $=0\farcs9$,
and the bottom panels are PSF FWHM $=1\farcs2$.
Red filled circles indicate the number counts measured from the simulated GREX-PLUS images
after correcting for both the output magnitude offset and detection completeness,
while red open circles indicate the number counts corrected only for the output magnitude offset
without the completeness correction.
Blue filled circles indicate the number counts measured from the JWST images of the same field;
no completeness correction is applied to the JWST-based number counts.
The vertical gray dashed lines mark the $5\sigma$ limiting aperture magnitude.
For the PSF FWHM $=0\farcs9$ case, the limiting magnitude is measured in circular apertures with a diameter of $2\farcs0$,
whereas for the PSF FWHM $=1\farcs2$ case, it is measured in circular apertures with a diameter of $2\farcs4$.
}
\label{fig:number_counts_corrected}
\end{center}
\end{figure*}

\subsection{First-order Integration Time Estimates for GREX-PLUS}
\label{subsec:First-order Integration Time Estimates for GREX-PLUS}

\hspace{0.8em}
To provide a first-order estimate, 
we calculate the GREX-PLUS integration time 
required to reach a given depth by scaling the JWST integration times 
used for the original NIRCam images.  
At wavelengths around $4\,\mu$m,
the total system throughput of JWST NIRCam
including the JWST optical telescope element,
the NIRCam optical train, dichroics, filters, and the detector quantum efficiency
is about $0.5$.\footnote{\url{https://jwst-docs.stsci.edu/jwst-near-infrared-camera/nircam-instrumentation/nircam-filters\#gsc.tab=0}}
This value is comparable to the expected throughput for GREX-PLUS
(\textcolor{blue}{GREX-PLUS Science Team et al. in preparation}).
Assuming that both observations are in the background-limited regime,
so that the noise is dominated by sky background rather than detector noise,
the difference in integration times between JWST and GREX-PLUS
can be primarily attributed to the difference in the telescope apertures. 
Specifically, the integration time required to reach a fixed limiting magnitude
scales approximately as the inverse square of the telescope diameter.
Here the limiting magnitudes are evaluated on a consistent basis 
using the same photometric aperture sizes. 
The JWST integration times for each field are summarized in
Section~\ref{sec:GREX-PLUS PSF and Ghost Model}.
We thus simply scale these integration times by a factor of $6.5^{2}$
to estimate the corresponding GREX-PLUS integration time.

Figure~\ref{fig:limitmag_texp} illustrates the $5\sigma$ limiting magnitudes 
measured from the simulated GREX-PLUS images
as a function of the estimated GREX-PLUS integration time.
Although the number of data points is limited,
a least-squares fit of the form 
$5\sigma$ (mag) $= A + B \log \left[ t_{\rm int} \, ({\rm hr}) \right]$ yields
$(A,\,B)=(25.65,\,0.73)$ for the case with PSF FWHM $=0.9$~arcsec
and $(A,\,B)=(25.41,\,0.63)$ for PSF FWHM $=1.2$~arcsec.
If the impact of source confusion is negligible,
the sensitivity is expected to improve as the square root of integration time,
which corresponds to a slope of $B=1.25$ in this relation.
Consistent with the limiting magnitude comparison discussed above,
our obtained slopes are smaller than this value.

Using these fitted relations, we estimate the integration time required to reach $26.5$~mag,
which corresponds to the target depth of the Deep layer introduced in 
Section~\ref{sec:GREX-PLUS PSF and Ghost Model}.
We obtain $t_{\rm int}\simeq 14.6$~hr for the case with PSF FWHM $=0.9$~arcsec 
and $t_{\rm int}\simeq 53.2$~hr for PSF FWHM $=1.2$~arcsec.
These values are roughly comparable to, or up to a factor of $\simeq 3$ larger than,
the required integration times estimated from the current GREX-PLUS baseline specifications 
described in Section~\ref{sec:GREX-PLUS PSF and Ghost Model}.
Although this calculation relies on a simplified scaling, 
our results highlight that a broader PSF could significantly increase 
the integration time required to reach the target depth of the Deep layer.
Mitigating this issue may require improvements such as
increasing the telescope aperture,
reducing pointing jitter through better telescope stability during observations, 
improving optical accuracy in construction and orbit,
and applying more advanced source extraction and image reconstruction methods
(e.g., iterative PSF-fitting, image deconvolution, or prior-based deblending 
utilizing high resolution ancillary data such as from the Roman Space Telescope).

\subsection{Completeness-corrected Number Counts}
\label{subsec:Completeness-corrected Number Counts}

\hspace{0.8em}
In Section~\ref{subsec:Source Extraction and Number Counts},
we have detected sources in the simulated GREX-PLUS images 
and derived the number counts as a function of the SExtractor output total magnitude.
In that analysis, we have used the output magnitudes as measured, 
i.e., without correcting for any systematic offset relative to the input magnitudes,
and the number counts have not yet been corrected for detection incompleteness.
Here, we consider the Monte Carlo simulation results in Section~\ref{subsec:Detection Completeness}.
We utilize the median offsets between the input and output total magnitudes
and the detection completeness as a function of total magnitude
to derive completeness-corrected number counts.

In Figure~\ref{fig:number_counts_corrected}, 
the red open circles show the number counts 
after correcting for the systematic magnitude offsets derived from the Monte Carlo simulations, 
while the red filled circles further include the correction for detection completeness.
In all panels, applying the Monte Carlo-based corrections extends the range 
over which the simulated GREX-PLUS counts follow the JWST-based counts, 
compared to the uncorrected results.
For the case with PSF FWHM $=0.9$~arcsec, 
the completeness-corrected GREX-PLUS number counts
are broadly consistent with the JWST-based number counts within the uncertainties down to $\simeq26$~mag.
Before applying these corrections, the GREX-PLUS-based number counts 
start to fall below the JWST-based number counts at around $24$~mag, 
indicating that detection incompleteness and measurement biases
have a significant impact on the faint end number counts.
Note that the faintest bin data points for GREX-PLUS 
still deviate from the JWST data points even after the completeness correction.
Given that the aperture correction relative to the PSF is 
about $0.4$~mag (Section \ref{sec:Simulated Imaging Data for GREX-PLUS}),
this residual offset is plausibly attributable to measurements being close to the detection limit,
where both photometric bias and completeness corrections become sensitive to the adopted extraction setup.
We therefore interpret the remaining mismatch in the faintest bin with caution.
Taken together with the limiting magnitude analysis in Section~\ref{subsec:Limiting Magnitude Comparison},
which shows that the depth estimated from blank sky aperture fluctuations continues to improve
over the range explored here, these results suggest that the simulated GREX-PLUS images
do not show a clear signature of having reached the confusion limit.

For the case with PSF FWHM $=1.2$~arcsec, the agreement is achieved over a brighter magnitude range:
the regime where the GREX-PLUS number counts remain broadly consistent with the JWST-based number counts
is approximately $0.5$~mag brighter than in the PSF FWHM $=0.9$~arcsec case,
while the overall behavior is otherwise similar.
Overall, these results indicate that statistical studies of faint galaxy populations,
such as number count measurements, should remain feasible for GREX-PLUS,
provided that source confusion effects are properly accounted for in survey planning and analysis.

\section{Summary}
\label{sec:summary}

\hspace{0.8em}
In this study, we have quantitatively investigated the impact of source confusion,
which can become a limiting factor in deep observations at long wavelengths, 
by adopting an empirical approach based on the latest GREX-PLUS optical model. 
We have generated simulated GREX-PLUS images near $4\,\mu$m 
with a range of effective integration times
by convolving the GREX-PLUS PSF and ghost kernel 
with deep, high angular resolution JWST NIRCam imaging data 
from extragalactic surveys of different depths: JADES, GLASS, CEERS, and COSMOS-Web.
For the GREX-PLUS PSF, 
we have taken into account the uncertainties from pointing jitter and optical system imperfections,
and considered two representative cases: an optimistic case with highly accurate optical performance
and a baseline case corresponding to the nominal specification, 
yielding PSF FWHM values of $0.9$~arcsec and $1.2$~arcsec, respectively.
For reference, in the current optical model, 
the total ghost flux is estimated to be about $4.6$\,\% of the nominal PSF flux,
suggesting that the ghost contribution is likely secondary 
relative to the extended PSF wings and the background fluctuations from unresolved sources.
Our main results are summarized as follows.

\begin{enumerate}

\item For both the original JWST NIRCam images and the simulated GREX-PLUS images,
we have estimated the limiting magnitudes 
by placing random circular apertures of the same sizes in regions with no detected sources 
and deriving the standard deviations from the flux distributions measured in those apertures. 
We have found that, 
for the moderately deep JWST datasets,  
the simulated GREX-PLUS limiting magnitudes are broadly consistent with the JWST-based values, 
whereas for the deeper JWST data the simulated GREX-PLUS images become systematically shallower, 
with the discrepancy being larger in the broader PSF case. 
This behavior likely reflects source confusion effects,
including confusion noise from the superposition of unresolved faint sources, 
as well as additional contributions from extended PSF wings and ghost halos 
that redistribute light from bright sources to larger radii, 
thereby elevating the background fluctuations measured in nominally blank regions.

\vspace{0.5em}

\item We have found that the limiting magnitudes of the simulated GREX-PLUS images 
continue to improve as the original JWST images become deeper, 
without showing a clear plateau over the depth range explored here.
This suggests that the simulated GREX-PLUS images do not yet exhibit a distinct confusion-limited plateau,
at least down to $5\sigma \simeq 27$--$27.5$~mag measured in $\simeq 2^{\prime\prime}$-diameter apertures
for the cases considered.
As a complementary view, we have also estimated the GREX-PLUS integration times 
corresponding to the underlying JWST datasets
and fitted the relation between the limiting magnitude and integration time. 
The best-fit slopes are shallower than the expectation for purely background-limited scaling,
indicating that the depth improvement becomes progressively less efficient toward longer integrations,
even though a clear plateau is not yet observed within the range probed here.
This effect should be taken into account when planning deep GREX-PLUS surveys.

\vspace{0.5em}

\item Using the GLASS and CEERS field data, 
which provide a combination of relatively wide area
and substantial depth among the datasets considered here, 
we have performed Monte Carlo simulations
by injecting point sources spanning a range of magnitudes and attempting to recover them.
From these simulations, we have quantified the detection completeness 
and the relation between input and output magnitudes, 
and used them to correct the observed number counts 
for systematic measurement bias and detection completeness.
After applying these corrections, the GREX-PLUS number counts become broadly consistent with
the JWST-based number counts over a total magnitude range down to around $26$~mag 
for PSF FWHM $=0.9$~arcsec.
For PSF FWHM $=1.2$~arcsec, a comparable level of agreement is obtained over a brighter range,
approximately $0.5$~mag shallower, while the overall behavior is otherwise similar.

\end{enumerate}

\vspace{0.5em}

Overall, our results indicate that source confusion effects already contribute 
at the depths relevant to the planned GREX-PLUS deep imaging, 
both through confusion noise from background fluctuations due to unresolved sources
and through confusion-induced blending that reduces completeness.
Although statistical studies of distant galaxy populations should remain feasible,
survey optimization should account for the progressively reduced efficiency of depth improvement toward longer integrations.
In particular, survey efficiency is sensitive to the effective PSF FWHM.
Reducing PSF broadening, for example by increasing the telescope aperture and 
reducing pointing jitter through better telescope stability,  
can enable reaching a given depth with shorter integration times.

\begin{ack}

\hspace{0.8em}
We thank Ken Mawatari and Nao Suzuki 
for providing insightful comments on our results. 
The authors acknowledge 
the CEERS team led by Steven L. Finkelstein, 
the GLASS team led by Tommaso Treu, 
the COSMOS-Web team led by Jeyhan Kartaltepe, 
and the JADES team led by Daniel Eisenstein and Nora Luetzgendorf 
for developing their observing programs.
This research made use of 
SExtractor (\citealt{1996A&AS..117..393B}), 
IRAF (\citealt{1986SPIE..627..733T,1993ASPC...52..173T}),\footnote{IRAF is distributed by the National Optical Astronomy Observatory, 
which is operated by the Association of Universities for Research in Astronomy (AURA) 
under a cooperative agreement with the National Science Foundation.} 
SAOImage DS9 \citep{2003ASPC..295..489J},
Numpy \citep{2020Natur.585..357H}, 
Matplotlib \citep{2007CSE.....9...90H}, 
Scipy \citep{2020NatMe..17..261V}, 
and Astropy \citep{2013A&A...558A..33A,2018AJ....156..123A}.\footnote{\url{http://www.astropy.org}}
%
This work was partially performed using the computer facilities of
the Institute for Cosmic Ray Research, The University of Tokyo. 
\end{ack}

\section*{Funding}

\hspace{0.8em}
This work was supported 
by 
KAKENHI Grant Numbers 
22K03670, 
23H00131, %
23H05441, %
23K17695, %
24H00245, 
25H00674, 
25K24561, %
26H02069, %
26K07135, 
and 26K17200 %
through the Japan Society for the Promotion of Science (JSPS). 
This work was partially supported by the joint research program of 
the Institute for Cosmic Ray Research (ICRR), University of Tokyo. 

\section*{Data availability} 

\hspace{0.8em}
This work is based in part on observations
made with the NASA/ESA/CSA James Webb Space Telescope. 
The data 
(\citealt{https://doi.org/10.17909/z7p0-8481}; 
\citealt{https://doi.org/10.17909/kw3c-n857}; 
\citealt{https://doi.org/10.17909/8tdj-8n28}) 
were obtained from the Mikulski Archive for Space Telescopes 
at the Space Telescope Science Institute, 
which is operated by the Association of Universities for Research in Astronomy, Inc., 
under NASA contract NAS 5-03127 for JWST.
These observations are associated with programs 
ERS-1345 (CEERS), 
ERS-1324 (GLASS), 
GO-1727 (COSMOS-Web), 
GTO-1180, GTO-1181, GTO-1210, GTO-1286, 
GO-1895, GO-1963, and GO-3215 (JADES).

\bibliographystyle{apj}
\bibliography{ref}

\end{document}